\documentclass[a4paper,11pt,dvips]{article}
\usepackage{graphicx}
\usepackage{amsmath}
\usepackage{amssymb}
\usepackage{latexsym}
\usepackage[colorlinks]{hyperref}
\usepackage{color}
\usepackage{float}
\usepackage{cite}
\usepackage{makeidx}
\usepackage{colortbl}
\newcommand{\bra}{\begin{array}}
\newcommand{\era}{\end{array}}
\newcommand{\beq}{\begin{equation}}
\newcommand{\eeq}{\end{equation}}
\newcommand{\bqr}{\begin{eqnarray}}
\newcommand{\eqr}{\end{eqnarray}}

\def\BC{\bb C}
\def\_\BC{\bbi C}

\def\( {\left(}
   \def\) {\right)}
\def\[ {\left[}
\def\] {\right]}
\def\no2 {{\textstyle{n\over 2}}}

\newcommand{\om}{\omega}

\newcommand{\ga}{\gamma}

\newcommand{\lga}{\longrightarrow}

\newcommand{\lb}{\label}

\begin{document}
\begin{titlepage}
\setcounter{page}{1}
\renewcommand{\thefootnote}{\fnsymbol{footnote}}

\begin{flushright}
%ucd-tpg 10-01\\
%arXiv:yymm.xxxx
\end{flushright}

\vspace{5mm}
\begin{center}

{\Large \bf {Path Integral for Confined Dirac Fermions in a Constant \\ Magnetic Field}}

\vspace{5mm}
{\bf Abdeldjalil Merdaci\footnote{\sf amerdaci@kfu.edu.sa}}$^{a,b}$,
 {\bf Ahmed Jellal\footnote{\sf ajellal@ictp.it --
a.jellal@ucd.ac.ma}}$^{c,d}$
 and {\bf Lyazid Chetouani}$^{b}$

\vspace{5mm}

{$^a$\em Physics Department, College of Science, King Faisal University,\\
 PO Box
380, Alahsa 31982, Saudi Arabia}

{$^b$\em D{\'{e}}partement de Physique, Facult\'{e} des Sciences Exactes,
Universit\'{e} Mentouri, \\ 25000 Constantine, Alg\'{e}rie.}

{$^c$\em Saudi Center for Theoretical Physics, Dhahran, Saudi Arabia}

{$^{d}$\em Theoretical Physics Group,  %Department of Physics,
Faculty of Sciences, Choua\"ib Doukkali University},\\
{\em PO Box 20, 24000 El Jadida, Morocco}

%{$^{d}$\em Max Planck Institute for the Physics of Complex Systems,\\
%N\"othnitzer Str. 38, 01187 Dresden} %\\[1em]

%\vspace{30mm}

\vspace{3cm}

\begin{abstract}
We consider Dirac fermion confined in harmonic potential and
submitted to a constant magnetic field. The corresponding
solutions of the energy spectrum are obtained by using the path
integral techniques. For this, we begin by establishing a symmetric
global projection, which provides a symmetric form for the Green function.
Based on this,
we show that it is possible
to end up with the propagator of the harmonic oscillator for
one charged particle. After some transformations, we derive the normalized
wave functions and the eigenvalues in terms of different physical parameters
and quantum numbers.
By interchanging quantum numbers, we show that our solutions have interesting properties.
The density of current and the non-relativistic limit are analyzed where different conclusions
are obtained. Finally, the
completeness of the Dirac oscillator eigenfunctions is proved by using
the standard properties of
the generalized Laguerre polynomials.

\vspace{3cm}



\end{abstract}
\end{center}
\end{titlepage}

%%%%%%%%%%%%%%%%%%%%%%%%%%%%%%%%%%%
\section{Introduction}
%%%%%%%%%%%%%%%%%%%%%%%%%%%%%%%%%%%

The Dirac equation in (2+1)-dimensions is an important tool not only from
mathematical point of view but also from the large applications in physics. In
particular, several condensed matter phenomena point out to the existence of a
(2 +1)-dimensional energy spectrum determined by the relativistic Dirac
equation\cite{1}. For very recent works, one may consult
references\cite{2,3,4,5,6,7} and for early works relevant to our subject we
cite\cite{8,9}. As example graphene\cite{10}, which is a single layer of
carbon atoms arranged into a planar honeycomb lattice. This system has
attracted a considerable attention from both experimental and theoretical
researchers since its experimental realization in 2004\cite{11}. This is
because of its unique and outstanding mechanical, electronic, optical, thermal
and chemical properties\cite{12}. Most of these marvelous properties are due
to the apparently relativistic-like nature of its carriers, electrons behave
as massless Dirac fermions in graphene systems. In fact starting from the
original tight-binding Hamiltonian describing graphene it has been shown
theoretically that the low-energy excitations of graphene appear to be
massless chiral Dirac fermions.

On the other hand, due to recent technological advances in nano-fabrication
there were a lot of interest in the study of low dimensional quantum systems
such as quantum wells, quantum wires and quantum dots\cite{13}. In particular,
there has been considerable amount of work in recent years on semiconductors
confined structures, which finds applications in electronic and optoelectronic
devices. These show the relevance of the confinements in physics and therefore
deserve to be analyzed in other systems like relativistic ones. Furthermore,
an applied magnetic field perpendicular to the heterostructure systems
quantizes the energy levels in the plane, drastically affecting the density of
states giving rise to the famous quantum Hall effect\cite{14}.

A confined relativistic system subjected to a constant magnetic field was
studied normal to the plane in\cite{15}. In fact, the exact solution of the
Dirac equation in (2 + 1)-dimensions was obtained depending on various
parameters. The solution space consists of positive and negative energy
solutions, each of which splits into two disconnected subspaces depending on
the sign of an azimuthal quantum number $l$ $=0,\pm1,\pm2$,... and whether the
cyclotron frequency is larger or smaller than the oscillator frequency. The
spinor wave function was written in terms of the associated Laguerre
polynomials. For negative $l$, the relativistic energy spectrum was found to
be in finitely degenerate due to the fact that it is dependent of $l$.

Motivated by different investigations on the Dirac fermions in
(2+1)-dimensions, we give an exact solution of a problem that has been studied
at various levels by researchers dealing with different physical phenomena.
More precisely, we return to the problem studied in\cite{15}and use another
techniques to determine the solutions of the energy spectrum as well dealing
with related issues. Then, we consider a relativistic particle subjected to an
external magnetic field as well as to a confining potential. By using the path
integral formalism and making different transformations, we show that the
causal Green function can be written in terms of the propagator for the
harmonic oscillator for a charged particle. This allowed us to obtain various
solutions, in terms of different physical parameters and quantum numbers, and
emphasis similarities to, and differences from, already published work
elsewhere\cite{15,16}.

Furthermore, we give discussions of our results based on different physical
settings. The full rich space of solutions suggested enabled us to carry out a
deeper analysis in relation to various physical quantities. For instance, we
show that our energy remains invariant under the changes of the quantum
numbers characterizing our system behavior, reveals a non trivial symmetry of
the problem. Also, we obtain, as expected in the absence of applied voltage, a
null current density for both directions in the Cartesian representation.
However, this is not the case in polar coordinate. In fact, we show that the
radial current vanishes whereas the angular component does not. It is
dependent on various physical parameters in the problem. Additionally, we
discuss the non-relativistic limit of the problem.

The paper is organized as follows. In section 2, we show that how one can use
the global projection rather than local one for (2 + 1)-dimensional Dirac
equation in electromagnetic field. In section 3, we give the theoretical
formulation of the problem where different changes are introduced to simplify
the process for obtaining the solutions of the energy spectrum. More
precisely, we will use the causal Green function as well as different
techniques to solve our problem. We determine the eigenvalues and eigenspinors
in terms of different physical parameters and quantum numbers in section 4. In
section 5, we analysis our results by showing that our system has hidden
symmetries, those can be used to deal with different issues. In section 6, we
calculate the density current in polar coordinate system. We show how to
recover the nonrelativistic limit from our results in section 7.
The proof of the completeness relation of the Dirac oscillator eigenfunctions
will be presented in section 8. We conclude
our findings in the final section.

%%%%%%%%%%%%%%%%%%%%%%%%%%%%%%%%%%%%%%%%%%%%%
\section{ Symmetric global projection}
%%%%%%%%%%%%%%%%%%%%%%%%%%%%%%%%%%%%%%%%%%%%%%%%

Before embarking on our problem and investigate its basic features, let us
establish some mathematical tool that needed to deal with different issues.
For this, we start by considering a relativistic charged particle in
electromagnetic field to show that the global projection giving rises to an
new symmetric form for the Green function. For such system the causal Green
function $S^{c}(\mathbf{x}_{b},\mathbf{x}_{a})$ satisfies the two Dirac
equations\cite{17}
%\begin{subequations}
\begin{eqnarray}
&& (\gamma^{\mu}\left(  i\partial_{b\mu}-eA_{\mu}\left(  \mathbf{x}_{b}\right)
\right)  -m)S^{c}(\mathbf{x}_{b},\mathbf{x}_{a})   =\delta^{3}%
(\mathbf{x}_{b}-\mathbf{x}_{a})\\
&& S^{c}(\mathbf{x}_{b},\mathbf{x}_{a})\left(  \gamma^{\mu}\left(
-i\overleftarrow{\partial}_{a\mu}-eA_{\mu}\left(  \mathbf{x}_{a}\right)
\right)  -m\right)     =\delta^{3}(\mathbf{x}_{b}-\mathbf{x}_{a})
\end{eqnarray}
where the matrices $\gamma^{\mu}$ are defined by the relations generating
Clifford algebra
%\end{subequations}
\begin{equation}
\left[  \gamma^{\mu},\gamma^{\nu}\right]  =-2i\sigma^{\mu\nu},\qquad\left\{
\gamma^{\mu},\gamma^{\nu}\right\}  =2\eta^{\mu\nu}.%
\end{equation}
Actually, we have $\eta^{\mu\nu}=\mbox{diag}\left(  1,-1,-1\right)  $, with
($\mu,\nu=0,1,2$) and $\gamma^{\mu}=\frac{i}{2}\epsilon^{\mu\nu\lambda}%
\gamma_{\nu}\gamma_{\lambda}$ or equivalently%
\begin{equation}
\gamma^{0}=i\gamma^{1}\gamma^{2},\qquad\gamma^{1}=-i\gamma^{0}\gamma^{2}%
,\qquad\gamma^{2}=i\gamma^{0}\gamma^{1}.
\end{equation}
Formally, $S^{c}(\mathbf{x}_{b},\mathbf{x}_{a})$ is the matrix element in the
coordinate space
\begin{equation}
S^{c}(\mathbf{x}_{b},\mathbf{x}_{a})=\langle\mathbf{x}_{b}\mid S^{c}%
\mid\mathbf{x}_{a}\rangle \lb{5555}
\end{equation}
of the inverse Dirac operator $O_{-}^{-1}$
\begin{equation}
S^{c}=(\gamma^{\mu}\left(  p_{\mu}-eA_{\mu}\right)  -m)^{-1}=O_{-}^{-1}%
=O_{+}\left(  O_{-}O_{+}\right)  ^{-1}\lb{6666}
\end{equation}
where the operators $O_{\pm}$ are given by
\begin{equation}\lb{opm}
O_{\pm}=\gamma^{\mu}\left(  p_{\mu}-eA_{\mu}\right)  \pm m.
\end{equation}

We recall that Alexandrou {\it et al.} \cite{18} have been described a massive Dirac
particle in external vector and scalar fields by making use of the asymmetric
form. In the present study, we consider the symmetric form to write the Dirac
propagator in order to straightforwardly derive the corresponding spinors.
Indeed, the local inverse Dirac operator \eqref{6666} can be written in two equivalent forms%
\begin{equation}
S^{c}=(\gamma^{\mu}\left(  p_{\mu}-eA_{\mu}\right)  -m)^{-1}=O_{-}^{-1}%
=O_{+}S_{g}^{c}=S_{g}^{c}O_{+}\lb{8}
\end{equation}
where the Green operator for global projection is
\begin{equation}
S_{g}^{c}=\left(  O_{-}O_{+}\right)  ^{-1}=\left(  O_{+}O_{-}\right)  ^{-1}%
\end{equation}
and the label $g$ stands for global. It is clear that the matrix element of
$S_{g}^{c}$ verifies the quadratic Dirac equation
\begin{equation}
O_{-}O_{+}S_{g}^{c}(\mathbf{x}_{b},\mathbf{x}_{a})=O_{+}O_{-}S_{g}%
^{c}(\mathbf{x}_{b},\mathbf{x}_{a})=\delta^{3}(\mathbf{x}_{b}-\mathbf{x}_{a}).
\end{equation}
Now let us insert the completeness relation $\int\mid z\rangle\langle z\mid
d^{3}z=\mathbb{I}$ \ in the matrix element \eqref{5555} and use \eqref{8} to end up with
\begin{equation}
S^{c}(\mathbf{x}_{b},\mathbf{x}_{a})=\int\langle x_{b}\mid O_{+}\mid
z\rangle\langle z\mid S_{g}^{c}\mid x_{a}\rangle d^{3}z=\int\langle x_{b}\mid
S_{g}^{c}\mid z\rangle\langle z\mid O_{+}\mid x_{a}\rangle d^{3}z
\end{equation}
which implies the equality
\begin{equation}
\left(  \gamma^{\mu}\overrightarrow{\pi}_{\mu}\left(  b\right)  +m\right)
S_{g}^{c}(\mathbf{x}_{b},\mathbf{x}_{a})=S_{g}^{c}(\mathbf{x}_{b}%
,\mathbf{x}_{a})\left(  \gamma^{\mu}\overleftarrow{\pi}_{\mu}\left(  a\right)
+m\right)
\end{equation}
where the momentum operators read as%
\begin{equation}
\overrightarrow{\pi}_{\mu}=i{\tfrac{\overrightarrow{\partial}}{\partial
x^{\mu}}}-eA_{\mu}\left(  x\right)  ,\quad\overleftarrow{\pi}_{\mu}%
=-i{\tfrac{\overleftarrow{\partial}}{\partial x^{\mu}}}-eA_{\mu}\left(
x\right).
\end{equation}
The right and left derivatives are defined by
\begin{equation}
\hat{A}\tfrac{\overrightarrow{\partial}}{\partial x}\hat{B}=\hat{A}\left(
\tfrac{\partial}{\partial x}\hat{B}\right)  ,\quad\hat{A}\tfrac{\overleftarrow
{\partial}}{\partial x}\hat{B}=\left(  \tfrac{\partial}{\partial x}\hat
{A}\right)  \hat{B}.%
\end{equation}
Therefore, we can deduce the following new symmetric form
\begin{equation}\lb{144}
S^{c}(\mathbf{x}_{b},\mathbf{x}_{a})=\tfrac{1}{2}\left(  \gamma^{\mu
}\overrightarrow{\pi}_{\mu}\left(  b\right)  +m\right)  S_{g}^{c}%
(\mathbf{x}_{b},\mathbf{x}_{a})+\tfrac{1}{2}S_{g}^{c}(\mathbf{x}%
_{b},\mathbf{x}_{a})\left(  \gamma^{\mu}\overleftarrow{\pi}_{\mu}\left(
a\right)  +m\right)
\end{equation}
which is an interesting result and can be generalized to any dimension. We
will see how it can be used to determine explicitly the solution of the energy spectrum.

%%%%%%%%%%%%%%%%%%%%%%%%%%%%%%%%%%%%%%%%%%%%%%%
\section{Application to confined system}
%%%%%%%%%%%%%%%%%%%%%%%%%%%%%%%%%%%%%%%%%%%%%%

{{red}To be specific, let us consider a confined system of $\left(  2+1\right)
$-dimensional Dirac fermions in a constant magnetic field $\overrightarrow{B}%
$.
This confinement can be realized by modifying the operators $O_\pm$, given in \eqref{opm},
%In this case, we can change the momentum
%\begin{equation}
by the non-minimal substitution such that
$\vec{p}\longrightarrow\vec{p}-im\omega\vec{r}\gamma^{0}$, where $\om$ is the oscillator frequency.
%\end{equation}
%by introducing the confining potential with the oscillator frequency $\omega$
%and considering minimal coupling. Combining all to write the Dirac operators
%as
By choosing the symmetric gauge the symmetric gauge $\vec{A}=\frac{B}{2}\left(  -y,x\right)  $, we can write \eqref{opm}
\beq
O_{\pm}=\gamma^{0}p_{0}-\gamma^{1}\left(  p_{x}-\frac{|e|B}{2}y -i m\om x \ga^0\right)  -\gamma^{2}\left(
p_{y}+\frac{|e|B}{2}x -i m\om y \ga^0\right)  \pm m
\eeq
and using $\ga^0= i\ga^1 \ga^2$ to obtain in the compact form 
\beq
O_{\pm}=\gamma^{0}p_{0}-\gamma^{1}\left(  p_{x}+Gy\right)  -\gamma^{2}\left(
p_{y}-Gx\right)  \pm m
\eeq
 where we
have set the parameter $G=m\left(  \omega-\omega_{c}\right)  $ and the
cyclotron frequency $\omega_{c}=\frac{\mid e\mid B}{2m}\neq\omega$.
It is
clearly seen that both frequencies $\om$ and $\om_c$ come from different sources but regrouped in he same
place, which is an unique property of (2 + 1)-dimensional space.}

Now, by
using Schwinger proper time representation, we can express $S_{g}%
^{c}(\mathbf{x}_{b},\mathbf{x}_{a})$ in an integral expressions form
\begin{equation}
S_{g}^{c}(\mathbf{x}_{b},\mathbf{x}_{a})=\langle\mathbf{x}_{b}\mid\left(
O_{+}O_{-}\right)  ^{-1}\mid\mathbf{x}_{a}\rangle=-\frac{i}{2}\int_{0}%
^{\infty}d\lambda\langle\mathbf{x}_{b}\mid\exp\left(  i\frac{\lambda}%
{2}\left(  \mathcal{H}+i\varepsilon\right)  \right)  \mid\mathbf{x}_{a}\rangle
\end{equation}
where the Hamiltonian $\mathcal{H}$ is given by
\begin{equation}
\mathcal{H}=O_{+}O_{-}=\hat{p}_{0}^{2}-\hat{p}_{x}^{2}-\hat{p}_{y}^{2}%
-G^{2}\left(  \hat{x}^{2}+\hat{y}^{2}\right)  -m^{2}+2G\left(  \hat{x}\hat
{p}_{y}-\hat{y}\hat{p}_{x}\right)  +2iG\gamma^{1}\gamma^{2}
\end{equation}
for $G\neq0$.
Similar to\cite{19}, we represent $S_{g}^{c}(\mathbf{x}_{b},\mathbf{x}_{a})$ via the path integral
\begin{align}
S_{g}^{c}(\mathbf{x}_{b},\mathbf{x}_{a})  &  =-\frac{i}{2}\exp\left(
i\gamma \cdot\frac{\partial_{l}}{\partial\theta}\right)  \int_{0}^{\infty}%
d\lambda\int D\mathbf{x}D\mathbf{p}\ \int\mathcal{D}\psi\\
&  \times\exp\left\{  i\int_{0}^{1}\left[  -\mathbf{\dot{x}}.\mathbf{p}%
+\frac{\lambda}{2}\left(  \mathbf{p}^{2}-m^{2}-G^{2}\mathbf{x}^{2}\right)
\right.  \right. \nonumber\\
&  \left.  \left.  \left.  +2G\mathbf{x}\times\mathbf{p}-8iG\psi^{1}\psi
^{2}-i\psi_{\mu}\dot{\psi}^{\mu}\right]  d\tau+\psi_{\mu}\left(  1\right)
\psi^{\mu}\left(  0\right)  \right\}  \right\vert _{\theta=0}\nonumber
\end{align}
where $\mathbf{x}=$ $\left(  t,x,y\right)  $ and $\left(  \psi,\theta\right)$
 refer respectively to even and odd variables. They  satisfy boundary
conditions
\begin{equation}
\mathbf{x}(0)=\mathbf{x}_{a},\qquad\mathbf{x}(1)=\mathbf{x}_{b},\qquad
\psi^{\mu}(0)+\psi^{\mu}(1)=\theta^{\mu}. \label{20}%
\end{equation}
The path integration measure
\begin{equation}
\mathcal{D}\psi=D\psi\left[  \int_{\psi(1)+\psi(0)=0}D\psi\exp\left(  \int
_{0}^{1}\psi_{\mu}\dot{\psi}^{\mu}d\tau\right)  \right]  ^{-1}%
\end{equation}
has been considered above. Note that, by integrating over the path $t$, we
can see that the momentums become constants, i.e. $p_{0}$=const.

To go further, let us simplify our formalism by changing the momentum
variables to define two others as
\begin{equation}
p_{x}=\bar{p}_{x}-Gy,\qquad p_{y}=\bar{p}_{y}+Gx
\end{equation}
which allow to write
\begin{align}
S_{g}^{c}(\mathbf{x}_{b},\mathbf{x}_{a}) &  =-\frac{i}{2}\exp\left(
i\gamma \cdot\frac{\partial_{l}}{\partial\theta}\right)  \int_{0}^{\infty}%
d\lambda\int_{-\infty}^{+\infty}\frac{dp_{0}}{2\pi}e^{-ip_{0}\left(
t_{b}-t_{a}\right)  +\frac{i\lambda}{2}\left(  p_{0}^{2}-m^{2}\right)  }\int
DxDyD\bar{p}_{x}D\bar{p}_{y}\nonumber\\
&  \times\int\mathcal{D}\psi\exp\left\{  i\int_{0}^{1}\left[  \bar{p}_{x}%
\dot{x}+\bar{p}_{y}\dot{y}-\frac{\lambda}{2}\bar{p}_{x}^{2}-\frac{\lambda}%
{2}\bar{p}_{y}^{2}\right.  \right.  \nonumber\\
&  +\left.  \left.  \left.  G\left(  x\dot{y}-y\dot{x}\right)  -4i\lambda
G\psi^{1}\psi^{2}-i\psi_{\mu}\dot{\psi}^{\mu}\right]  d\tau+\psi_{\mu}\left(
1\right)  \psi^{\mu}\left(  0\right)  \right\}  \right\vert _{\theta=0}.%
\end{align}
Now integrating over $\bar{p}_{x}$ and $\bar{p}_{y}$ to find
\begin{align}
S_{g}(\mathbf{x}_{b},\mathbf{x}_{a}) &  =-\frac{i}{2}\exp\left(  i\gamma
.\frac{\partial_{l}}{\partial\theta}\right)  \int_{0}^{\infty}d\lambda
\int_{-\infty}^{+\infty}\frac{dp_{0}}{2\pi}e^{-ip_{0}\left(  t_{b}%
-t_{a}\right)  +\frac{i\lambda}{2}\left(  p_{0}^{2}-m^{2}\right)  }\int
DxDy\nonumber\label{24}\\
&  \times\int\mathcal{D}\psi e^{\left.  \left\{  i\int_{0}^{1}\left(
\frac{\dot{x}^{2}}{2\lambda}+\frac{\dot{y}^{2}}{2\lambda}+G\left(  x\dot
{y}-y\dot{x}\right)  -4i\lambda G\psi^{1}\psi^{2}-i\psi_{\mu}\dot{\psi}^{\mu
}\right)  d\tau+\psi_{\mu}\left(  1\right)  \psi^{\mu}\left(  0\right)
\right\}  \right\vert _{\theta=0}}.%
\end{align}

In order to integrate over $\psi^{\mu}\left(  \tau\right)  $, let us first
eliminate the constants $\theta^{\mu}$ appearing in the boundary conditions
\eqref{20} by performing the following changes%
\begin{equation}
\psi^{\mu}\left(  \tau\right)  =\xi^{\mu}\left(  \tau\right)  +\frac{1}%
{2}\theta^{\mu},\qquad\mu=0,1,2.
\end{equation}
By replacing in \eqref{24}, we obtain
\begin{align}
S_{g}^{c}(\mathbf{x}_{b},\mathbf{x}_{a}) &  =-\frac{i}{2}\exp\left(
i\gamma\cdot\frac{\partial_{l}}{\partial\theta}\right)  \int_{0}^{\infty}%
d\lambda\int_{-\infty}^{+\infty}\frac{dp_{0}}{2\pi}e^{-ip_{0}\left(
t_{b}-t_{a}\right)  +\frac{i\lambda}{2}\left(  p_{0}^{2}-m^{2}\right)  }\int
DxDy\nonumber\label{26}\\
&  \times\int\mathcal{D}\xi\exp\left\{  i\int_{0}^{1}\left[  \frac{\dot{x}%
^{2}}{2\lambda}+\frac{\dot{y}^{2}}{2\lambda}+G\left(  x\dot{y}-y\dot
{x}\right)  \right.  \right\}  \\
&  \left.  \left.  \left.  -4i\lambda G\left(  \xi^{1}\xi^{2}-\tfrac{1}%
{2}\theta^{2}\xi^{1}\left(  \tau\right)  +\tfrac{1}{2}\theta^{1}\xi^{2}\left(
\tau\right)  +\tfrac{1}{4}\theta^{1}\theta^{2}\right)  -i\xi_{\mu}\dot{\xi
}^{\mu}\right]  d\tau\right\}  \right\vert _{\theta=0}\nonumber
\end{align}
where the antiperiodic boundary conditions become
\begin{equation}
\xi^{\mu}(1)+\xi^{\mu}(0)=0.
\end{equation}
By considering $\mathbf{\xi=}\left(
\begin{array}
[c]{c}%
\xi^{1}\\
\xi^{2}%
\end{array}
\right)  $ and $\sigma_{2}=\left(
\begin{array}
[c]{cc}%
0 & -i\\
i & 0
\end{array}
\right)  $, one can easily see the relation
\begin{equation}
\xi^{1}\xi^{2}=\tfrac{i}{2}\mathbf{\xi}\sigma_{2}\mathbf{\xi}\label{28}%
\end{equation}
Now using \eqref{28} and being able to undertake easily the integration
over Grassmann variables, one can separate \eqref{26} as
\begin{align}
S_{g}^{c}(\mathbf{x}_{b},\mathbf{x}_{a}) &  =-\frac{i}{2}\exp\left(
i\gamma\cdot \frac{\partial_{l}}{\partial\theta}\right)  \int_{0}^{\infty}%
d\lambda\int_{-\infty}^{+\infty}\frac{dp_{0}}{2\pi}e^{-ip_{0}\left(
t_{b}-t_{a}\right)  +\frac{i\lambda}{2}\left(  p_{0}^{2}-m^{2}\right)
}\label{29}\\
&  \times\iint DxDy\exp\left\{  i\int_{0}^{1}\left(  \frac{\dot{x}^{2}%
}{2\lambda}+\frac{\dot{y}^{2}}{2\lambda}+G\left(  x\dot{y}-y\dot{x}\right)
\right)  \right\}  \left.  \mathcal{I}\left(  \lambda,\theta\right)
\right\vert _{\theta=0}\nonumber
\end{align}
where the fermionic part is given by
\begin{align}
\mathcal{I}\left(  \lambda,\theta\right)   &  =e^{\lambda G\theta^{1}%
\theta^{2}}\int\mathcal{D}\xi^{0}\iint\mathcal{D}^{2}\mathbf{\xi}\\
&  \times e^{\int_{0}^{1}\mathbf{\xi}\left(  \tau\right)  \mathcal{R}\left(
\lambda\mid\tau,\tau^{\prime}\right)  \mathbf{\xi}\left(  \tau^{\prime
}\right)  d\tau d\tau^{\prime}+\int_{0}^{1}\mathcal{J}\left(  \tau\right)
\mathbf{\xi}\left(  \tau\right)  d\tau+\int_{0}^{1}\xi^{0}\left(  \tau\right)
\delta^{\prime}\left(  \tau-\tau^{\prime}\right)  \xi^{0}\left(  \tau^{\prime
}\right)  d\tau d\tau^{\prime}}\nonumber
\end{align}
and we have
\begin{equation}
\mathcal{R}\left(  \lambda\mid\tau,\tau^{\prime}\right)  \mathbf{=}%
-\delta^{\prime}\left(  \tau-\tau^{\prime}\right)  +2i\lambda G\sigma
_{2}\delta\left(  \tau-\tau^{\prime}\right)  ,\qquad\mathcal{J}^{T}\left(
\tau\right)  =2\lambda G\left(  -\theta^{2},\theta^{1}\right).
\end{equation}
Since the integration over $\xi^{0}$ yields trivially the unity, then integrating
over $\xi^{1}$ and $\xi^{2}$ to obtain
\begin{equation}
\mathcal{I}\left(  x,y,\lambda;\theta\right)  =\sqrt{\tfrac{\det
\mathcal{R}\left(  \lambda\right)  }{\det\mathcal{R}\left(  0\right)  }%
}e^{\lambda G\theta^{1}\theta^{2}}e^{\tfrac{1}{4}\int_{0}^{1}\int_{0}%
^{1}\mathcal{J}\left(  \tau\right)  .\mathcal{R}^{-1}\left(  \lambda\mid
\tau,\tau^{\prime}\right)  .\mathcal{J}\left(  \tau^{\prime}\right)  d\tau
d\tau^{\prime}}%
\end{equation}
where $\mathcal{R}^{-1}\left(  \lambda\mid\tau,\tau^{\prime}\right)  $ is the
inverse matrix of $\mathcal{R}\left(  \lambda\mid\tau,\tau^{\prime}\right)  $,
which can be considered as an operator acting on the space of antiperiodic
functions
\begin{equation}
\mathcal{R}^{-1}\left(  \lambda\mid1,\tau\right)  +\mathcal{R}^{-1}\left(
\lambda\mid0,\tau\right)  =0,\qquad\forall\tau\in\left[  0,1\right].
\end{equation}
It obeys the first order differential equation
\begin{equation}
\left(  -\frac{\partial}{\partial\tau}+2iG\lambda\sigma_{2}\right)
\mathcal{R}^{-1}\left(  \lambda\mid\tau,\tau^{\prime}\right)  =\delta\left(
\tau-\tau^{\prime}\right)
\end{equation}
giving the solution
\begin{equation}
\mathcal{R}^{-1}\left(  \lambda\mid\tau,\tau^{\prime}\right)  =\frac{1}%
{2}e^{2i\lambda G\mathbf{\sigma}_{2}\left(  \tau-\tau^{\prime}\right)
}\left[  i\sigma_{2}\tan\left(  \lambda G\right)  -\varepsilon\left(
\tau-\tau^{\prime}\right)  \right]  ,\quad\varepsilon\left(  \tau-\tau
^{\prime}\right)  =\mbox{sgn}\left(  \tau-\tau^{\prime}\right).
\end{equation}
Using similar calculations as in \cite{20} to get
\begin{eqnarray}
\sqrt{\tfrac{\det\mathcal{R}\left(  \lambda\right)  }{\det\mathcal{R}\left(
0\right)  }} &=& e^{\frac{1}{2}Tr\int_{0}^{\lambda}\mathcal{R}^{-1}\left(
\lambda^{\prime}\right)  \frac{d}{d\lambda^{\prime}}\mathcal{R}\left(
\lambda^{\prime}\right)  d\lambda^{\prime}}=e^{\frac{1}{2}Tr\int_{0}^{\lambda
}\left[  \frac{1}{2}i\sigma_{2}\tan\left(  \lambda^{\prime}G\right)  \right]
\left[  2iG\sigma_{2}\right]  d\lambda^{\prime}} \nonumber\\
&=& e^{-\int_{0}^{\lambda}%
\tan\left(  \lambda^{\prime}G\right)  Gd\lambda^{\prime}}=\cos\left(  \lambda
G\right).
\end{eqnarray}
Combining all to show that \eqref{29} takes the form
\begin{equation}\label{366}
S_{g}^{c}(\mathbf{x}_{b},\mathbf{x}_{a})=-\tfrac{i}{2}\int_{0}^{\infty
}d\lambda\int_{-\infty}^{+\infty}\frac{dp_{0}}{2\pi}e^{-ip_{0}\left(
t_{b}-t_{a}\right)  +\frac{i\lambda}{2}\left(  p_{0}^{2}-m^{2}\right)  }%
\Phi\left(  \lambda\right)  K\left(  x_{a},x_{b},y_{a},y_{b};\lambda\right)
\end{equation}
where $K$ is the propagator of the harmonic oscillator for one charged
particle
\begin{equation}
K\left(  x_{a},x_{b},y_{a},y_{b};\lambda\right)  =\iint DxDye^{i\int_{0}%
^{1}\left(  \frac{\dot{x}^{2}}{2\lambda}+\frac{\dot{y}^{2}}{2\lambda}+G\left(
x\dot{y}-y\dot{x}\right)  \right)  d\tau}%
\end{equation}
and the function $\Phi\left(  \lambda\right)  $ reads as
\begin{equation}
\Phi\left(  \lambda\right)  =\cos\left(  \lambda G\right)  \exp\left(
i\gamma\cdot\frac{\partial_{l}}{\partial\theta}\right)  \left.  \exp\left(
\tan\left(  \lambda G\right)  \theta^{1}\theta^{2}\right)  \right\vert
_{\theta=0}. %
\end{equation}
In order to evaluate the propagator $K$, we first decouple $x$ and $y$ by
introducing a rotation of coordinates to define new variables $q_{1}$ and
$q_{2}$ as
\begin{align}
&  x=q_{1}\cos\left(  \lambda G\tau\right)  +\kappa q_{2}\sin\left(  \lambda
G\tau\right)  \\
&  y=-\kappa q_{1}\sin\left(  \lambda G\tau\right)  +q_{2}\cos\left(  \lambda
G\tau\right)
\end{align}
where $\kappa$ $=\mbox{sgn}\left(  G\right)  \neq0$. Under the above rotation,
$K$ can be written in separable form as
\begin{equation}
K\left(  x_{a},x_{b},y_{a},y_{b};\lambda\right)  =\iint Dq_{1}Dq_{2}%
e^{i\int_{0}^{1}\left(  \frac{\dot{q}_{1}^{2}}{2\lambda}+\frac{\dot{q}_{2}%
^{2}}{2\lambda}-\frac{\lambda\left\vert G\right\vert ^{2}}{2}q_{1}^{2}%
-\frac{\lambda\left\vert G\right\vert ^{2}}{2}q_{2}^{2}\right)  d\tau}%
\end{equation}
which represents the kernel of two propagators corresponding to two harmonic
oscillators and has well-known form%
\begin{equation}
K\left(  x_{a},x_{b},y_{a},y_{b};\lambda\right)  =\tfrac{\left\vert
G\right\vert }{2i\pi\sin\left(  \left\vert G\right\vert \lambda\right)
}e^{\frac{i\left\vert G\right\vert }{2\sin\left\vert G\right\vert \lambda
}\left[  \left(  \left(  x_{b}-x_{a}\right)  ^{2}+\left(  y_{b}-y_{a}\right)
^{2}\right)  \cos\left(  \left\vert G\right\vert \lambda\right)  +2\kappa
\sin\left(  \left\vert G\right\vert \lambda\right)  \left(  x_{a}y_{b}%
-y_{a}x_{b}\right)  \right]  }%
\end{equation}
where the following coordinates have been used
\begin{align}
&  q_{1a}=x_{a},\qquad q_{1b}=\cos\left(  G\lambda\right)  x_{b}-\kappa
\sin\left(  G\lambda\right)  y_{b}\\
&  q_{2a}=y_{a},\qquad q_{2b}=\kappa\sin\left(  G\lambda\right)  x_{b}%
+\cos\left(  G\lambda\right)  y_{b}. %
\end{align}
To explicitly determine the matrix $\Phi\left(  \lambda\right)  $, let us
proceed by using derivation over the variables $\theta$. The calculations
performed by adopting the prevailing approach \cite{19} such that acting the operator
$\tfrac{\partial_{l}}{\partial\theta}$ and then replacing $\theta$ by the
matrices $\gamma$. We can write the expression of $\Phi\left(  \lambda\right)
$ by using the identities%
\begin{eqnarray}
&&\exp\left(  i\gamma^{\mu}\tfrac{\partial_{l}}{\partial\theta^{\mu}}\right)
f(\theta)\mid_{\theta=0}=f\left(  \tfrac{\partial_{l}}{\partial\zeta}\right)
\exp(i\zeta^{\mu}\gamma_{\mu})\mid_{\zeta=0},
\\
&& \exp\left(  i\zeta^{\mu}\gamma_{\mu}\right)  =1+i\zeta^{\mu}\gamma_{\mu}%
+\frac{1}{2}\zeta^{\mu}\zeta^{\nu}\gamma_{\mu}\gamma_{\nu}+i\zeta^{0}\zeta
^{1}\zeta^{2}\gamma_{0}\gamma_{1}\gamma_{2}
\end{eqnarray}
which are valid for (2+1)-dimensional, with
\begin{equation}
\gamma^{1}=i\sigma_{1},\qquad\gamma^{2}=i\sigma_{2},\qquad\gamma^{0}%
=\sigma_{3}%
\end{equation}
where $\zeta^{\mu}$ are odd variables. In this case, we can simplify
$\Phi(\lambda)$ as%
\begin{equation}
\Phi\left(  \lambda\right)  =e^{i\kappa\left\vert G\right\vert \lambda
\sigma_{3}}%
\end{equation}
and therefore \eqref{366} becomes
\begin{equation}
S_{g}^{c}(\mathbf{x}_{b},\mathbf{x}_{a})=-\tfrac{i}{2}\int_{0}^{\infty
}d\lambda\int_{-\infty}^{+\infty}\frac{dp_{0}}{2\pi}e^{-ip_{0}\left(
t_{b}-t_{a}\right)  +\frac{i\lambda}{2}\left(  p_{0}^{2}-m^{2}\right)
}e^{i\kappa\left\vert G\right\vert \lambda\sigma_{3}}K\left(  x_{a}%
,x_{b},y_{a},y_{b};\lambda\right)
\end{equation}
which can be implemented in \eqref{144} to obtain the causal Green
function. This task will be done to extract the solutions of the energy spectrum.

%%%%%%%%%%%%%%%%%%%%%%%%%%%%%%%%
\section{Energy spectrum}
%%%%%%%%%%%%%%%%%%%%%%%%%%%%%%%%%%

We would like to determine the eigenvalues and corresponding eigenspinors of
our system.To this end, we introduce the polar coordinates $x=r\cos
\varphi,y=r\sin\varphi.$ Thus, the causal Green function \eqref{144}
becomes
\begin{align}
S^{c}\left(r_{a},r_{b},\varphi_{a},\varphi_{b}, t_b,t_a\right)   &  =-\frac{i}{4}%
\int_{0}^{\infty}d\lambda\int_{-\infty}^{+\infty}\frac{dp_{0}}{2\pi}%
e^{-ip_{0}\left(  t_{b}-t_{a}\right)  +\frac{i\lambda}{2}\left(  p_{0}%
^{2}-m^{2}\right)  } \lb{500}\\
&  \times\left\{  \left(  p_{0}\sigma_{3}+\sigma_{1}e^{i\sigma_{3}\varphi_{b}%
}\tfrac{\partial}{\partial r_{b}}+\frac{1}{r_{b}}\sigma_{2}e^{i\sigma
_{3}\varphi_{b}}\tfrac{\partial}{\partial\varphi_{b}}-i\kappa\left\vert
G\right\vert \sigma_{2}r_{b}e^{i\sigma_{3}\varphi_{b}}+m\right)
e^{i\kappa\left\vert G\right\vert \lambda\sigma_{3}}K\right.  \nonumber\\
&  \left.  +e^{i\kappa\left\vert G\right\vert \lambda\sigma_{3}}\left(
p_{0}\sigma_{3}-\sigma_{1}e^{i\sigma_{3}\varphi_{a}}\tfrac{\partial}{\partial
r_{a}}-\frac{1}{r_{a}}\sigma_{2}e^{i\sigma_{3}\varphi_{a}}\tfrac{\partial
}{\partial\varphi_{a}}-i\kappa\left\vert G\right\vert \sigma_{2}%
r_{a}e^{i\sigma_{3}\varphi_{a}}+m\right)  K\right\}  \nonumber
\end{align}
where
\begin{equation}
K=K\left(  r_{a},r_{b},\varphi_{a},\varphi_{b};\lambda\right)  =\tfrac
{\left\vert G\right\vert }{2i\pi\sin\left(  \left\vert G\right\vert
\lambda\right)  }e^{\frac{i\left\vert G\right\vert }{2\sin\left\vert
G\right\vert \lambda}\left[  \left(  r_{b}^{2}+r_{a}^{2}\right)  \cos\left(
\left\vert G\right\vert \lambda\right)  -2r_{b}r_{a}\left(  \cos\left(
\varphi_{b}-\varphi_{a}+\kappa\left\vert G\right\vert \lambda\right)  \right)
\right]  }.%
\end{equation}
In order to deduce the wave functions, we use the expansion and Hille formulas
\cite{22}
\begin{align}
&  e^{\left(  -\frac{i\left\vert G\right\vert }{\sin\left\vert G\right\vert
\lambda}r_{b}r_{a}\cos\left(  \varphi_{b}-\varphi_{a}+\kappa\left\vert
G\right\vert \lambda\right)  \right)  }=\sum_{l=-\infty}^{+\infty
}I_{\left\vert l\right\vert }\left(  -\frac{i\left\vert G\right\vert }%
{\sin\left\vert G\right\vert \lambda}r_{b}r_{a}\right)  e^{il\left(
\varphi_{b}-\varphi_{a}+\kappa\left\vert G\right\vert \lambda\right)  }\\
&  \tfrac{\left(  xyt\right)  ^{-\frac{\alpha}{2}}}{1-t}\exp\left(
-t\tfrac{x+y}{1-t}\right)  I_{\alpha}\left(  \tfrac{2\sqrt{xyt}}{1-t}\right)
=\sum_{n=0}^{\infty}n!\tfrac{L_{n}^{\alpha}\left(  x\right)  L_{n}^{\alpha
}\left(  y\right)  t^{n}}{\Gamma\left(  n+\alpha+1\right)  },\qquad\left\vert
t\right\vert <1.
\end{align}
By taking $t=e^{-2i\lambda\left\vert G\right\vert },x=\left\vert G\right\vert
r_{b}^{2},y=\left\vert G\right\vert r_{a}^{2}$ and $\alpha=\left\vert
l\right\vert ,$ the propagator $K$ can be expressed as
\begin{eqnarray}
K\left(  r_{a},r_{b},\varphi_{a},\varphi_{b};\lambda\right)   &
= &\tfrac{\left\vert G\right\vert }{\pi}\sum_{n=0}^{+\infty}\sum_{l=-\infty
}^{+\infty}e^{-2i\lambda\left\vert G\right\vert n}e^{-i\left\vert l\right\vert
\left\vert G\right\vert \lambda}e^{il\left(  \varphi_{b}-\varphi_{a}%
+\kappa\left\vert G\right\vert \lambda\right)  }\tfrac{n!}{\Gamma\left(
n+\left\vert l\right\vert +1\right)  }\nonumber\label{55}\\
&&  \times e^{-i\left\vert G\right\vert \lambda}z_{b}^{\left\vert l\right\vert
}z_{a}^{\left\vert l\right\vert }e^{-\frac{1}{2}z_{b}^{2}}e^{-\frac{1}{2}%
z_{a}^{2}}L_{n}^{\left\vert l\right\vert }\left(  z_{b}^{2}\right)
L_{n}^{\left\vert l\right\vert }\left(  z_{a}^{2}\right)
\end{eqnarray}
where $z=\sqrt{\left\vert G\right\vert }r$, $L_{n}^{\left\vert l\right\vert
}\left(  z^{2}\right)  $ is a generalized Laguerre polynomial and $\left\vert
l\right\vert $ is an integer number. After inserting \eqref{55}
into \eqref{500} we obtain
\begin{eqnarray}
S^{c}(\mathbf{x}_{b},\mathbf{x}_{a}) &  =&-\tfrac{i\left\vert G\right\vert
\sqrt{\left\vert G\right\vert }}{4\pi}\sum_{n=0}^{+\infty}\sum_{l=-\infty
}^{+\infty}\tfrac{n!}{\left(  n+\left\vert l\right\vert \right)  !}%
\int_{-\infty}^{+\infty}\frac{dp_{0}}{2\pi}e^{-ip_{0}\left(  t_{b}%
-t_{a}\right)  }e^{il\left(  \varphi_{b}-\varphi_{a}\right)  }\nonumber\\
&&  \times\int_{0}^{\infty}d\lambda e^{\frac{i}{2}\lambda\left(  p_{0}%
^{2}-m^{2}-2\left\vert G\right\vert \left[  2n+1+\left\vert l\right\vert
-\kappa l\right]  \right)  }\nonumber \lb{555}\\
&&  \times\left\{  \left(  \left(  -i\kappa\sigma_{2}z_{b}+\sigma_{1}%
\tfrac{\partial}{\partial z_{b}}+i\sigma_{2}\tfrac{l}{z_{b}}\right)
e^{i\sigma_{3}\varphi_{b}}+\tfrac{m+p_{0}\sigma_{3}}{\sqrt{\left\vert
G\right\vert }}\right)  e^{i\kappa\left\vert G\right\vert \lambda\sigma_{3}%
}\right.  \nonumber\\
&&  +\left.  e^{i\kappa\left\vert G\right\vert \lambda\sigma_{3}}\left(
e^{-i\sigma_{3}\varphi_{a}}\left(  -i\kappa\sigma_{2}z_{a}-\sigma_{1}%
\tfrac{\partial}{\partial z_{a}}+i\sigma_{2}\tfrac{l}{z_{a}}\right)
+\tfrac{m+p_{0}\sigma_{3}}{\sqrt{\left\vert G\right\vert }}\right)  \right\}
\nonumber\\
&&  \times e^{-\frac{1}{2}z_{b}^{2}}z_{b}^{\left\vert l\right\vert }%
L_{n}^{\left\vert l\right\vert }\left(  z_{b}^{2}\right)  e^{-\frac{1}{2}%
z_{a}^{2}}z_{a}^{\left\vert l\right\vert }L_{n}^{\left\vert l\right\vert
}\left(  z_{a}^{2}\right).
\end{eqnarray}
Let us introduce the spin operator $\sigma_{3}$ and insert the identity
$\sum_{s=\pm1}\chi_{s}\chi_{s}^{+}=\mathbb{I}$, where $\sigma_{3}\chi_{s}=s\chi_{s}$
and $\chi_{s}^{+}\sigma_{3}=s\chi_{s}^{+}$. We can easily check the relations
\begin{align}
&  \sigma_{1}\chi_{s}=\chi_{-s},\qquad\sigma_{2}\chi_{s}=is\chi_{-s}\\
&  \chi_{s}^{+}\sigma_{1}=\chi_{-s}^{+},\qquad\chi_{s}^{+}\sigma_{2}%
=-is\chi_{-s}^{+}%
\end{align}
for the vectors
\begin{equation}
\chi_{+1}=\left(
\begin{array}
[c]{c}%
1\\
0
\end{array}
\right)  ,\qquad\chi_{-1}=\left(
\begin{array}
[c]{c}%
0\\
1
\end{array}
\right)
\end{equation}
and the identities
\begin{equation}
\sigma_{2}e^{i\sigma_{3}\varphi}=e^{-i\sigma_{3}\varphi}\sigma_{2},\quad
\sigma_{1}e^{i\sigma_{3}\varphi}=e^{-i\sigma_{3}\varphi}\sigma_{1}.%
\end{equation}
These can be used to write the Green function \eqref{555} relative to
our particle as
\begin{eqnarray}
S^{c}(\mathbf{x}_{b},\mathbf{x}_{a}) &  =&-\tfrac{i\left\vert G\right\vert
\sqrt{\left\vert G\right\vert }}{4\pi}\sum_{n=0}^{+\infty}\sum_{l=-\infty
}^{+\infty}\sum_{s=\pm1}\tfrac{n!}{\left(  n+\left\vert l\right\vert \right)
!}\int_{-\infty}^{+\infty}\frac{dp_{0}}{2\pi}e^{-ip_{0}\left(  t_{b}%
-t_{a}\right)  }e^{il\left(  \varphi_{b}-\varphi_{a}\right)  }\nonumber\\
&&  \times\int_{0}^{\infty}d\lambda e^{\frac{i}{2}\lambda\left(  p_{0}%
^{2}-m^{2}-2\left\vert G\right\vert \left[  2n+1+\left\vert l\right\vert
-\kappa\left(  l+s\right)  \right]  \right)  }\nonumber\\
&&  \times\left\{  \left(  \left(  -i\kappa\sigma_{2}z_{b}+\sigma_{1}%
\tfrac{\partial}{\partial z_{b}}+i\sigma_{2}\tfrac{l}{z_{b}}\right)
e^{is\varphi_{b}}+\tfrac{m+p_{0}s}{\sqrt{\left\vert G\right\vert }}\right)
\chi_{s}\chi_{s}^{+}e^{i\kappa\left\vert G\right\vert \lambda s}\right.
\nonumber\\
&&  +\left.  e^{i\kappa\left\vert G\right\vert \lambda s}\chi_{s}\chi_{s}%
^{+}\left(  e^{-is\varphi_{a}}\left(  -i\kappa\sigma_{2}z_{a}-\sigma_{1}%
\tfrac{\partial}{\partial z_{a}}+i\sigma_{2}\tfrac{l}{z_{a}}\right)
+\tfrac{m+p_{0}s}{\sqrt{\left\vert G\right\vert }}\right)  \right\}
\nonumber\\
& & \times e^{-\frac{1}{2}z_{b}^{2}}z_{b}^{\left\vert l\right\vert }%
L_{n}^{\left\vert l\right\vert }\left(  z_{b}^{2}\right)  e^{-\frac{1}{2}%
z_{a}^{2}}z_{a}^{\left\vert l\right\vert }L_{n}^{\left\vert l\right\vert
}\left(  z_{a}^{2}\right).
\end{eqnarray}
In order to symmetrize the angular part we make the shift
\begin{equation}
l\rightarrow l-\frac{s+1}{2}%
\end{equation}
which implies%
\begin{eqnarray}
  S^{c}(\mathbf{x}_{b},\mathbf{x}_{a}) &=&-\tfrac{i\left\vert G\right\vert
\sqrt{\left\vert G\right\vert }}{4\pi}\sum_{n=0}^{\infty}\sum_{l=-\infty
}^{+\infty}\sum_{s=\pm1}\tfrac{\left\vert n\right\vert !}{\left(  \left\vert
n\right\vert +\left\vert l-\frac{s+1}{2}\right\vert \right)  !}\int_{-\infty
}^{+\infty}\frac{dp_{0}}{2\pi}e^{-ip_{0}\left(  t_{b}-t_{a}\right)
}e^{il\left(  \varphi_{b}-\varphi_{a}\right)  }e^{-i\frac{\varphi_{b}%
-\varphi_{a}}{2}}\nonumber\\
& & \times\int_{0}^{\infty}d\lambda e^{\frac{i}{2}\lambda\left(  p_{0}%
^{2}-m^{2}-2\left\vert G\right\vert \left[  2n+1+\left\vert l-\frac{s+1}%
{2}\right\vert -\kappa\left(  l+\frac{s-1}{2}\right)  \right]  \right)
}\nonumber\\
&&  \times\left\{  \left(  \left(  \kappa sz_{b}+\tfrac{\partial}{\partial
z_{b}}-s\tfrac{l-\frac{s+1}{2}}{z_{b}}\right)  \chi_{-s}\chi_{s}^{+}%
e^{is\frac{\varphi_{b}+\varphi_{a}}{2}}+\tfrac{m+p_{0}s}{\sqrt{\left\vert
G\right\vert }}e^{-is\frac{\varphi_{b}-\varphi_{a}}{2}}\chi_{s}\chi_{s}%
^{+}\right)  \right.  \nonumber\\
&&  +\left.  \left(  \left(  -\kappa sz_{a}-\tfrac{\partial}{\partial z_{a}%
}+s\tfrac{l-\frac{s+1}{2}}{z_{a}}\right)  \chi_{s}\chi_{-s}^{+}e^{-is\frac
{\varphi_{b}+\varphi_{a}}{2}}+\chi_{s}\chi_{s}^{+}\tfrac{m+p_{0}s}%
{\sqrt{\left\vert G\right\vert }}e^{-is\frac{\varphi_{b}-\varphi_{a}}{2}%
}\right)  \right\}  \nonumber\\
&&  \times e^{-\frac{1}{2}z_{b}^{2}}z_{b}^{\left\vert l-\frac{s+1}%
{2}\right\vert }L_{n}^{\left\vert l-\frac{s+1}{2}\right\vert }\left(
z_{b}^{2}\right)  e^{-\frac{1}{2}z_{a}^{2}}z_{a}^{\left\vert l-\frac{s+1}%
{2}\right\vert }L_{n}^{\left\vert l-\frac{s+1}{2}\right\vert }\left(
z_{a}^{2}\right).
\end{eqnarray}
Introducing
\begin{equation}
F_{n,l,s}^{\kappa^{\prime}}\left(  z\right)  =\sqrt{\tfrac{\left(  n\right)
!}{\left(  n+\kappa^{\prime}\left(  l-\frac{s+1}{2}\right)  \right)  !}%
}e^{-\frac{1}{2}z^{2}}z^{\kappa^{\prime}\left(  l-\frac{s+1}{2}\right)  }%
L_{n}^{\kappa^{\prime}\left(  l-\frac{s+1}{2}\right)  }\left(  z^{2}\right)
\end{equation}
where $\kappa^{\prime}=\mbox{sgn}\left(  l-\frac{s+1}{2}\right)  $ and the
momentum
\begin{equation}
\pi_{s}^{\kappa}\left(  z\right)  =\frac{d}{dz}+\kappa sz-s\frac{l-\frac
{s+1}{2}}{z}%
\end{equation}
we write $S^{c}(\mathbf{x}_{b},\mathbf{x}_{a})$ as
\begin{eqnarray}
S^{c}(\mathbf{x}_{b},\mathbf{x}_{a}) &  =& -\tfrac{i\left\vert G\right\vert
\sqrt{\left\vert G\right\vert }}{4\pi}\sum_{n=0}^{+\infty}\sum_{l=-\infty
}^{+\infty}\sum_{s=\pm1}\int_{-\infty}^{+\infty}\frac{dp_{0}}{2\pi}%
e^{-ip_{0}\left(  t_{b}-t_{a}\right)  }e^{-\frac{1}{2}z_{b}^{2}}e^{-\frac
{1}{2}z_{a}^{2}}e^{il\left(  \varphi_{b}-\varphi_{a}\right)  }e^{-i\frac
{\varphi_{b}-\varphi_{a}}{2}}\nonumber\\
&&  \times\sum_{s=\pm1}\int_{0}^{\infty}d\lambda e^{-\frac{i}{2}\lambda\left(
p_{0}^{2}-m^{2}-2\left\vert G\right\vert \left[  2n+1+\left\vert l-\frac
{s+1}{2}\right\vert -\kappa\left(  l+\frac{s-1}{2}\right)  \right]  \right)
}\\
& & \times\left\{  \left(  e^{is\frac{\varphi_{b}+\varphi_{a}}{2}}\pi
_{s}^{\kappa}\left(  z_{b}\right)  \chi_{-s}\chi_{s}^{+}+e^{-is\frac
{\varphi_{b}-\varphi_{a}}{2}}\tfrac{m+p_{0}s}{\sqrt{\left\vert G\right\vert }%
}\chi_{s}\chi_{s}^{+}\right)  F_{n,l,s}^{\kappa^{\prime}}\left(  z_{b}\right)
F_{n,l,s}^{\kappa^{\prime}}\left(  z_{a}\right)  \right.  \nonumber\\
&&  +\left.  \left(  -e^{-is\frac{\varphi_{b}+\varphi_{a}}{2}}\pi_{s}^{\kappa
}\left(  z_{a}\right)  \chi_{s}\chi_{-s}^{+}+e^{-is\frac{\varphi_{b}%
-\varphi_{a}}{2}}\tfrac{m+p_{0}s}{\sqrt{\left\vert G\right\vert }}\chi_{s}%
\chi_{s}^{+}\right)  F_{n,l,s}^{\kappa^{\prime}}\left(  z_{b}\right)
F_{n,l,s}^{\kappa^{\prime}}\left(  z_{a}\right)  \right\}\nonumber.
\end{eqnarray}
Integrating over $\lambda$ to get the expression
\beq
\left[  \frac
{i}{2}\left(  p_{0}^{2}-m^{2}-2\left\vert G\right\vert \left[  2n+1+\left\vert
l-\frac{s+1}{2}\right\vert -\kappa\left(  l+\frac{s-1}{2}\right)  \right]
\right)  \right]  ^{-1}
\eeq
which has the poles
\begin{equation}
p_{0}=\pm\sqrt{m^{2}+2\left\vert G\right\vert \left[  2n+1+\left\vert
l-\tfrac{s+1}{2}\right\vert -\kappa\left(  l+\tfrac{s-1}{2}\right)  \right]
}=\pm E_{n,l,s}.%
\end{equation}
After using the residue theorem at pole $p_{0}$, we find
\begin{equation}
\int_{-\infty}^{+\infty}f\left(  p_{0}\right)  \frac{dp_{0}}{2\pi}%
\frac{e^{-ip_{0}\left(  t_{b}-t_{a}\right)  }}{p_{0}^{2}-E_{n,l,s}^{2}}=-i%
%TCIMACRO{\dsum \limits_{\varepsilon=\pm1}}%
%BeginExpansion
{\displaystyle\sum\limits_{\varepsilon=\pm1}}
%EndExpansion
f\left(  \varepsilon E_{n,l,s}\right)  \frac{e^{-i\varepsilon E_{n,l,s}\left(
t_{b}-t_{a}\right)  }}{2E_{n,l,s}}\Theta\left(  \varepsilon\left(  t_{b}%
-t_{a}\right)  \right)
\end{equation}
where $\Theta(x)$ is the Heaviside step function.
Since $\varepsilon=\pm1$ and $s=\pm1$, one can easily check the identity %on peut verifier facillement l'identite
%suivante
$\sum_{s=\pm1}f_{s}\sum_{\varepsilon=\pm1}g_{\varepsilon}=\sum
_{s=\pm1}f_{s}\left(  g_{s}+g_{-s}\right).$
After
rearranging different terms, we obtain
\begin{align}
S^{c}(\mathbf{x}_{b},\mathbf{x}_{a}) &  =\frac{i}{2\pi}\sqrt{\left\vert
G\right\vert ^{3}}\sum_{n=0}^{+\infty}\sum_{l=-\infty}^{+\infty}\sum_{s=\pm
1}\frac{1}{2E_{n,l,s}}e^{i(l-1/2)\left(  \varphi_{b}-\varphi_{a}\right)
}\nonumber\\
&  \times\left\{  {e^{-isE_{n,l,s}\left(  t_{b}-t_{a}\right)  }}\Theta\left(
s\left(  t_{b}-t_{a}\right)  \right)  \left[  \left(  e^{is\frac{\varphi
_{b}+\varphi_{a}}{2}}\pi_{s}^{\kappa}\left(  z_{b}\right)  \chi_{-s}\chi
_{s}^{+}+e^{-is\frac{\varphi_{b}-\varphi_{a}}{2}}\tfrac{m+E_{n,l,s}}%
{\sqrt{\left\vert G\right\vert }}\chi_{s}\chi_{s}^{+}\right)  \right.
\right.  \nonumber\\
&  \left.  -\left(  e^{-is\frac{\varphi_{b}+\varphi_{a}}{2}}\pi_{s}^{\kappa
}\left(  z_{a}\right)  \chi_{s}\chi_{-s}^{+}-e^{-is\frac{\varphi_{b}%
-\varphi_{a}}{2}}\tfrac{m+E_{n,l,s}}{\sqrt{\left\vert G\right\vert }}\chi
_{s}\chi_{s}^{+}\right)  \right]   \lb{68}\\
&  \left.  +{e^{isE_{n,l,s}\left(  t_{b}-t_{a}\right)  }}\Theta\left(
-s\left(  t_{b}-t_{a}\right)  \right)  \left[  \left(  e^{is\frac{\varphi
_{b}+\varphi_{a}}{2}}\pi_{s}^{\kappa}\left(  z_{b}\right)  \chi_{-s}\chi
_{s}^{+}+e^{-is\frac{\varphi_{b}-\varphi_{a}}{2}}\tfrac{m-E_{n,l,s}}%
{\sqrt{\left\vert G\right\vert }}\chi_{s}\chi_{s}^{+}\right)  \right.
\right.  \nonumber\\
&  \left.  \left.  -\left(  e^{-is\frac{\varphi_{b}+\varphi_{a}}{2}}\pi
_{s}^{\kappa}\left(  z_{a}\right)  \chi_{s}\chi_{-s}^{+}-e^{-is\frac
{\varphi_{b}-\varphi_{a}}{2}}\tfrac{m-E_{n,l,s}}{\sqrt{\left\vert G\right\vert
}}\chi_{s}\chi_{s}^{+}\right)  \right]  \right\}  F_{n,l,s}^{\kappa^{\prime}%
}\left(  z_{b}\right)  F_{n,l,s}^{\kappa^{\prime}}\left(  z_{a}\right)\nonumber.
\end{align}
To go further, we proceed by using
%By implementing
the following mapping
\begin{equation}\lb{69}
s\rightarrow s^{\prime}=-s,\qquad n\rightarrow n^{\prime}=n-\frac
{\kappa^{\prime}+\kappa}{2}s=\left\{
\begin{array}
[l]{lll}%
n-\kappa s, & \text{ if }\kappa^{\prime}=\kappa\text{ or }\kappa^{\prime}=0\text{
}\\
n,& \text{ if }\kappa^{\prime}=-\kappa
\end{array}
\right.
\end{equation}
only for terms containing $\Theta\left(  -s\left(  t_{b}-t_{a}\right)
\right)  $ in \eqref{68} to the dummy variables $n$ and $s$, which
%Note that \eqref{69}
does not alter the integer nature of $n$. By
taking into consideration the energy invariance under \eqref{69}, the
Green function takes the following form%
\begin{eqnarray}
S^{c}(\mathbf{x}_{b},\mathbf{x}_{a}) &  =&\frac{i}{2\pi}\sqrt{\left\vert
G\right\vert ^{3}}\sum_{n=0}^{+\infty}\sum_{l=-\infty}^{+\infty}\sum_{s=\pm
1}\frac{1}{2E_{n,l,s}}e^{i(l-1/2)\left(  \varphi_{b}-\varphi_{a}\right)
}{e^{-isE_{n,l,s}\left(  t_{b}-t_{a}\right)  }}\Theta\left(  s\left(
t_{b}-t_{a}\right)  \right)  \nonumber\\
& & \times\left\{  \left[  \left(  e^{is\frac{\varphi_{b}+\varphi_{a}}{2}}%
\pi_{s}^{\kappa}\left(  z_{b}\right)  \chi_{s^{\prime}}\chi_{s}^{+}%
+e^{-is\frac{\varphi_{b}-\varphi_{a}}{2}}\tfrac{m+E_{n,l,s}}{\sqrt{\left\vert
G\right\vert }}\chi_{s}\chi_{s}^{+}\right)  \right.  \right.  \\
& & \left.  +\left(  -e^{-is\frac{\varphi_{b}+\varphi_{a}}{2}}\pi_{s}^{\kappa
}\left(  z_{a}\right)  \chi_{s}\chi_{s^{\prime}}^{+}+e^{-is\frac{\varphi
_{b}-\varphi_{a}}{2}}\tfrac{m+E_{n,l,s}}{\sqrt{\left\vert G\right\vert }}%
\chi_{s}\chi_{s}^{+}\right)  \right]  F_{n,l,s}^{\kappa^{\prime}}\left(
z_{b}\right)  F_{n,l,s}^{\kappa^{\prime}}\left(  z_{a}\right)  \nonumber\\
&&  +\left[  \left(  e^{-is\frac{\varphi_{b}+\varphi_{a}}{2}}\pi_{s^{\prime}%
}^{\kappa}\left(  z_{b}\right)  \chi_{s}\chi_{s^{\prime}}^{+}+e^{is\frac
{\varphi_{b}-\varphi_{a}}{2}}\tfrac{m-E_{n,l,s}}{\sqrt{\left\vert G\right\vert
}}\chi_{s^{\prime}}\chi_{s^{\prime}}^{+}\right)  \right.  \nonumber\\
& & \left.  \left.  +\left(  -e^{is\frac{\varphi_{b}+\varphi_{a}}{2}}%
\pi_{s^{\prime}}^{\kappa}\left(  z_{a}\right)  \chi_{s^{\prime}}\chi_{s}%
^{+}+e^{is\frac{\varphi_{b}-\varphi_{a}}{2}}\tfrac{m-E_{n,l,s}}{\sqrt
{\left\vert G\right\vert }}\chi_{s^{\prime}}\chi_{s^{\prime}}^{+}\right)
\right]  F_{n^{\prime},l,s^{\prime}}^{\kappa^{\prime}}\left(  z_{b}\right)
F_{n^{\prime},l,s^{\prime}}^{\kappa^{\prime}}\left(  z_{a}\right)  \right\}
\nonumber.
\end{eqnarray}
By virtue of Rodrigues' formula%
\begin{equation}
L_{n}^{\alpha}\left(  x\right)  =\frac{1}{n!}e^{x}x^{-\alpha}\frac{d^{n}%
}{dx^{n}}\left(  e^{-x}x^{n+\alpha}\right)
\end{equation}
we deduce after verification for all numbers $(\kappa$, $\kappa^{\prime}$,
$s$) the interesting properties
\begin{align}
&  \pi_{s}^{\kappa}\left(  z\right)  F_{n,l,s}^{\kappa^{\prime}}\left(
z\right)  =-s\sqrt{\tfrac{E_{n,l,s}^{2}-m^{2}}{\left\vert G\right\vert }%
}F_{n^{\prime},l,s^{\prime}}^{\kappa^{\prime}}\left(  z\right)  \\
&  \pi_{s^{\prime}}^{\kappa}\left(  z\right)  F_{n^{\prime},l,s^{\prime}%
}^{\kappa^{\prime}}\left(  z\right)  =s\sqrt{\tfrac{E_{n,l,s}^{2}-m^{2}%
}{\left\vert G\right\vert }}F_{n,l,s}^{\kappa^{\prime}}\left(  z\right).
\end{align}
These give
\begin{eqnarray}
S^{c}(\mathbf{x}_{b},\mathbf{x}_{a}) &  =& i\tfrac{\left\vert G\right\vert }%
{\pi}\sum_{n=0}^{+\infty}\sum_{l=-\infty}^{+\infty}\sum_{s=\pm1}\frac
{1}{2E_{n,l,s}}e^{i(l-1/2)\left(  \varphi_{b}-\varphi_{a}\right)
}e^{-isE_{n,l,s}\left(  t_{b}-t_{a}\right)  }\Theta\left(  s\left(
t_{b}-t_{a}\right)  \right)  \nonumber\\
&&  \times\left\{  e^{-is\frac{\varphi_{b}-\varphi_{a}}{2}}({E_{n,l,s}%
+m})F_{n,l,s}^{\kappa^{\prime}}\left(  z_{b}\right)  F_{n,l,s}^{\kappa
^{\prime}}\left(  z_{a}\right)  \chi_{s}\chi_{s}^{+}\right.  \nonumber\\
&&  +e^{is\frac{\varphi_{b}-\varphi_{a}}{2}}({E_{n,l,s}-m})F_{n^{\prime
},l,s^{\prime}}^{\kappa^{\prime}}\left(  z_{b}\right)  F_{n^{\prime
},l,s^{\prime}}^{\kappa^{\prime}}\left(  z_{a}\right)  \chi_{s^{\prime}}%
\chi_{s^{\prime}}^{+}\nonumber\\
&&  -s\sqrt{E_{n,l,s}^{2}-m^{2}}\left[  e^{-is\frac{\varphi_{b}+\varphi_{a}}%
{2}}F_{n,l,s}^{\kappa^{\prime}}\left(  z_{b}\right)  F_{n^{\prime}%
,l,s^{\prime}}^{\kappa^{\prime}}\left(  z_{a}\right)  \chi_{s}\chi_{s^{\prime
}}^{+}\right.  \nonumber\\
&&  \left.  \left.  +e^{is\frac{\varphi_{b}+\varphi_{a}}{2}}F_{n^{\prime
},l,s^{\prime}}^{\kappa^{\prime}}\left(  z_{b}\right)  F_{n,l,s}%
^{\kappa^{\prime}}\left(  z_{a}\right)  \chi_{s^{\prime}}\chi_{s}^{+}\right]
\right\}  s\sigma_{3}%
\end{eqnarray}
which can be simplified to end up with
\begin{equation}
S^{c}(\mathbf{x}_{b},\mathbf{x}_{a})=i\sum_{n=0}^{+\infty}\sum_{l=-\infty
}^{+\infty}\sum_{s=\pm1}\Psi_{n,l,s}^{\kappa,\kappa^{\prime}}\left(
r_{b},\varphi_{b};t_{b}\right)  \left(  \Psi_{n,l,s}^{\kappa,\kappa^{\prime}%
}\left(  r_{a},\varphi_{a};t_{a}\right)  \right)  ^{+}\sigma_{3}s\Theta\left(
s\left(  t_{b}-t_{a}\right)  \right)
\end{equation}
where the normalized wave functions are given by
\begin{equation}
\Psi_{n,l,s}^{\kappa,\kappa^{\prime}}\left(  r,\varphi;t\right)  =\sqrt
{\tfrac{\left\vert G\right\vert }{\pi}}e^{-i\mathcal{E}_{n,l,s}t}e^{il\varphi
}e^{-i\frac{\sigma_{3}+1}{2}\varphi}\left[  \sqrt{\tfrac{\mathcal{E}%
_{n,l,s}+sm}{2\mathcal{E}_{n,l,s}}}F_{n,l,s}^{\kappa^{\prime}}\left(
z\right)  \chi_{s}-s\sqrt{\tfrac{\mathcal{E}_{n,l,s}-sm}{2\mathcal{E}_{n,l,s}%
}}F_{n-\frac{\kappa^{\prime}+\kappa}{2}s,l,-s}^{\kappa^{\prime}}\left(
z\right)  \chi_{-s}\right]
\end{equation}
and the corresponding eigenvalues take the form
\begin{equation}
\mathcal{E}_{n,l,s}=sE_{n,l,s}=s\sqrt{m^{2}+2\left\vert G\right\vert \left[
2n+1+\kappa^{\prime}\left(  l-\tfrac{s+1}{2}\right)  -\kappa\left(
l+\tfrac{s-1}{2}\right)  \right]  }.%
\end{equation}
In compact form, we have the solutions of the energy spectrum
\begin{eqnarray}
\Psi_{n,l,s}\left(  r,\varphi;t\right)  &=& \sqrt{\tfrac{\left\vert G\right\vert
}{\pi}}e^{-is\mathcal{E}_{n,l,s}t}e^{il\varphi}e^{-i\frac{\sigma_{3}+1}%
{2}\varphi} \\
&& \times\left[  \sqrt{\tfrac{\mathcal{E}_{n,l,s}+sm}{2\mathcal{E}_{n,l,s}}%
}F_{n,l,s}\left(  z\right)  \chi_{s}+\sqrt{\tfrac{\left\vert G\right\vert
}{2\mathcal{E}_{n,l,s}\left(  \mathcal{E}_{n,l,s}+sm\right)  }}\pi_{s}\left(
z\right)  F_{n,l,s}\left(  z\right)  \chi_{-s}\right]\nonumber
\end{eqnarray}
\begin{equation}\lb{8822}
\mathcal{E}_{n,l,s}=s\sqrt{m^{2}+2\left\vert G\right\vert \left[
2n+1+\left\vert l-\tfrac{s+1}{2}\right\vert -\tfrac{G}{\left\vert G\right\vert
}\left(  l+\tfrac{s-1}{2}\right)  \right]  }%
\end{equation}
with the quantities $\kappa=\mbox{sgn}\left(  G\right)  =\frac{G}{\left\vert
G\right\vert }\neq0$, $\kappa^{\prime}=\mbox{sgn}\left(  l-\tfrac{s+1}%
{2}\right)  ,s=\mbox{sgn}\left(  \mathcal{E}_{n,l,s}\right)  $ and the
variable $z=\sqrt{\left\vert G\right\vert }r$ . It is interesting to note that
our obtained energy spectrum is completely in agreement with that derived
quantum mechanically in \cite{15}

Note that by putting $\om= 0$, which gives $G = -m\om_c$, and making the substitution
$l\lga l+\frac{s+1}{2}$ in \eqref{8822} we end up with the energy
\beq\lb{8833}
E^2_{s=\pm 1} = m^2 + 2m\om_c \left[2n + 1 + |l| + l + s\right].
\eeq
This can be compared to the energy for a massive fermion in a constant magnetic field in 3+1-dimensional space-time \cite{23}.
It is given by
\beq
E^2_{\lambda=\pm 1} - p^2_z
= m^2 + 2m\kappa \left[2n + |l| + l -\lambda + 1\right]
\eeq
which shows that it is in agreement with our derived result \eqref{8833}.

%%%%%%%%%%%%%%%%%%%%%%%%%%%%%%%%%%%%%%%%%
\section{Hidden symmetries}
%%%%%%%%%%%%%%%%%%%%%%%%%%%%%%%%%%%%%%%

We show that there are nontrivial hidden symmetries in our solutions
of the energy spectrum. As far as the eigenvalues are concerned, we notice that by
considering three configurations of quantum number $\left(  n,l,s\right)  $,
the energy remains invariant absolutely. These are listed in the following table
\begin{center}%
\begin{tabular}
[c]{|c|c|c|c|}\hline\hline
Sign $s$ & Quantum number $l$ & Quantum number $n$ & Energy $\mathcal{E}%
_{n,l,s}$\\\hline\hline
$s\longrightarrow-s$ & $\quad l\longrightarrow l-s$ & $n\longrightarrow
n-\kappa s$ & $\mathcal{E}_{n-\kappa s,l-s,-s}=-\mathcal{E}_{n,l,s}%
$\\\hline\hline
$s\longrightarrow-s$ & $l\longrightarrow l+s$ & $n\longrightarrow
n-\kappa^{\prime}s$ & $\mathcal{E}_{n-\kappa^{\prime}s,l+s,-s}=-\mathcal{E}%
_{n,l,s}$\\\hline\hline
$s\longrightarrow-s$ & $l\longrightarrow l$ & $n\longrightarrow n-\frac
{\kappa^{\prime}+\kappa}{2}s$ & $\mathcal{E}_{n-\frac{\kappa^{\prime}+\kappa
}{2}s,l,-s}=-\mathcal{E}_{n,l,s}$\\\hline\hline
\end{tabular}
\end{center}

Table 1: {\sf Table summarizes different symmetries of our system according
to changes in terms of the quantum numbers} $\left(  n,l,s\right)  $.\\

Furthermore, similar attention reed to be paid to the eigenspinors
$\Psi_{n,l,s}\left(  r,\varphi;t\right)  $ and underline their basic features.
Let us focus particularly on the last configuration and express $\Psi
_{n,l,s}\left(  r,\varphi;t\right)  $ as
\begin{equation}
\Psi_{n,l,s}\left(  r,\varphi;t\right)  =u_{n,l,s}\left(  r,\varphi;t\right)
\chi_{s}+v_{n,l,s}\left(  r,\varphi;t\right)  \chi_{-s}%
\end{equation}
where $u_{n,l,s}$ and $v_{n,l,s}$ are two-component defined by%
\begin{align}
u_{n,l,s}\left(  r,\varphi;t\right)   &  =e^{-i\mathcal{E}_{n,l,s}t}%
\sqrt{\tfrac{\left\vert G\right\vert }{\pi}}e^{il\varphi}e^{-i\frac{s+1}%
{2}\varphi}\sqrt{\tfrac{\mathcal{E}_{n,l,s}+sm}{2\mathcal{E}_{n,l,s}}%
}F_{n,l,s}^{\kappa^{\prime}}\left(  z\right) \\
v_{n,l,s}\left(  r,\varphi;t\right)   &  =-se^{-i\mathcal{E}_{n,l,s}t}%
\sqrt{\tfrac{\left\vert G\right\vert }{\pi}}e^{il\varphi}e^{-i\frac{-s+1}%
{2}\varphi}\sqrt{\tfrac{\mathcal{E}_{n,l,s}-sm}{2\mathcal{E}_{n,l,s}}%
}F_{n-\frac{\kappa^{\prime}+\kappa}{2}s,l,-s}^{\kappa^{\prime}}\left(
z\right).
\end{align}
Now using the third configuration in Table 1 to show there is a relation
between $v_{n,l,s}$ and $u_{n,l,s}$. This is given by
\begin{equation}
v_{n-\frac{\kappa^{\prime}+\kappa}{2}s,l,-s}\left(  r,\varphi;-t\right)
=s\sqrt{\frac{\mathcal{E}_{n,l,s}-sm}{\mathcal{E}_{n,l,s}+sm}}u_{n,l,s}\left(
r,\varphi;t\right)
\end{equation}
which is in agreement with \cite{233}. According to the above results, we
conclude that the obtained energy spectrum is invariant with respect to the
symmetries listed  above. This invariance can be used to deal with some issues
related to the present system.

%%%%%%%%%%%%%%%%%%%%%%%%%%%%%%%%%%%%%
\section{Density of current}
%%%%%%%%%%%%%%%%%%%%%%%%%%%%%%%%%%%%%

Let us consider the density of current $\vec{J}$ for our system and
investigate its basic features. Indeed, from our results we can end up with
the form%
\begin{equation}
\vec{J}=i\left\langle \sigma_{3}\vec{\sigma}\right\rangle .
\end{equation}
For this calculation, we use the spinor wavefunction obtained above. This
gives a null value in Cartesian coordinates, i.e. $J_{x}$ $=$ $J_{y}=0$, which
of course, is expected since there is no net charge drift in our system.

As a reassuring exercise, we calculate the same current $\vec{J}$ \ but in
cylindrical coordinates. Doing this process to obtain two components%
\beq
J_{r}=\vec{J}.\hat{r}=i\left\langle \sigma_{3}\vec{\sigma} \cdot\hat{r}%
\right\rangle ,\quad J_{\theta}=\vec{J} \cdot\hat{\theta}=i\left\langle \sigma
_{3}\vec{\sigma}.\hat{\theta}\right\rangle.
\eeq
We can show that the radial component is null, i.e. $\bar{J}_{r}^{n,l,s}=0$.
For the angular one, we use the definition to obtain
\begin{equation}
\bar{J}_{\varphi}^{n,l,s}=\int_{0}^{+\infty}\int_{0}^{2\pi}\Psi_{n,l,s}%
^{+}\left(  r,\varphi\right)  \sigma_{1}e^{i\sigma_{3}\varphi}\Psi
_{n,l,s}\left(  r,\varphi\right)  rdrd\varphi
\end{equation}
which can be written as
\begin{align}
\bar{J}_{\varphi}^{n,l,s} &  =2\left\vert G\right\vert \sqrt{\tfrac{\left\vert
G\right\vert }{\mathcal{E}_{n,l,s}^{2}}}\int_{0}^{+\infty}F_{n,l,s}\left(
z\right)  \pi_{s}^{\kappa}\left(  z\right)  F_{n,l,s}\left(  z\right)  rdr \label{87}\\
&  =2\left\vert G\right\vert \sqrt{\tfrac{\left\vert G\right\vert
}{\mathcal{E}_{n,l,s}^{2}}}\int_{0}^{+\infty}F_{n,l,s}\left(  z\right)
\left(  \frac{d}{dz}+\kappa sz-s\tfrac{l-\frac{s+1}{2}}{z}\right)
F_{n,l,s}\left(  z\right)  rdr\nonumber.
\end{align}
Using the relation \cite{21}
\begin{equation}
\int_{0}^{+\infty}x^{\alpha-1}e^{-x}L_{m}^{\gamma}\left(  x\right)
L_{n}^{\lambda}\left(  x\right)  dx=\tfrac{\left(  1+\gamma\right)
_{m}\left(  \lambda-\alpha+1\right)  _{n}\Gamma\left(  \alpha\right)  }%
{m!n!}_{3}F_{2}\left(  -m,\alpha,\alpha-\lambda;1+\gamma,\alpha-\lambda
-n;1\right)
\end{equation}
for $\operatorname{Re}\left(  \alpha\right)  >0,$ the well-known recurrence and
derivative properties of generalized Laguerre functions to obtain the final form
for \eqref{87}
\begin{eqnarray}\label{cccc}
\bar{J}_{\varphi}^{n,l,s} &  =& 2s\tfrac{n!\sqrt{\left\vert G\right\vert }%
}{\left(  n+\left\vert l-\frac{s+1}{2}\right\vert \right)  !\mathcal{E}%
_{n,l,s}}\left(  \left\vert l-\tfrac{s+1}{2}\right\vert -s\left(
l-\tfrac{s+1}{2}\right)  \right)  \tfrac{\left(  1+\left\vert l-\frac{s+1}%
{2}\right\vert \right)  _{n}\left(  \tfrac{1}{2}\right)  _{n}\Gamma\left(
\left\vert l-\tfrac{s+1}{2}\right\vert +\tfrac{1}{2}\right)  }{\left(
n!\right)  ^{2}}\nonumber\\
&&  \times_{3}F_{2}\left(  -n,\left\vert l-\tfrac{s+1}{2}\right\vert +\tfrac
{1}{2},\tfrac{1}{2};1+\left\vert l-\tfrac{s+1}{2}\right\vert ,\tfrac{1}%
{2}-n;1\right) \nonumber \\
&&  -4s\tfrac{n!\sqrt{\left\vert G\right\vert }}{\left(  n+\left\vert
l-\frac{s+1}{2}\right\vert \right)  !\mathcal{E}_{n,l,s}}\tfrac{\left(
2+\left\vert l-\frac{s+1}{2}\right\vert \right)  _{n-1}\left(  -\tfrac{1}%
{2}\right)  _{n}\Gamma\left(  \left\vert l-\tfrac{s+1}{2}\right\vert
+\tfrac{3}{2}\right)  }{\left(  n-1\right)  !n!}\nonumber\\
&&  \times_{3}F_{2}\left(  -n+1,\left\vert l-\tfrac{s+1}{2}\right\vert
+\tfrac{3}{2},\tfrac{3}{2};2+\left\vert l-\frac{s+1}{2}\right\vert ,\tfrac
{3}{2}-n;1\right)  \\
&&  +2s\tfrac{n!\sqrt{\left\vert G\right\vert }}{\left(  n+\left\vert
l-\frac{s+1}{2}\right\vert \right)  !\mathcal{E}_{n,l,s}}\left(  \tfrac
{G}{\left\vert G\right\vert }s-1\right)  \tfrac{\left(  1+\left\vert
l-\frac{s+1}{2}\right\vert \right)  _{n}\left(  -\tfrac{1}{2}\right)
_{n}\Gamma\left(  \left\vert l-\tfrac{s+1}{2}\right\vert +\tfrac{3}{2}\right)
}{\left(  n!\right)  ^{2}}\nonumber\\
&&  \times_{3}F_{2}\left(  -n,\left\vert l-\tfrac{s+1}{2}\right\vert +\tfrac
{3}{2},\tfrac{3}{2};1+\left\vert l-\tfrac{s+1}{2}\right\vert ,\tfrac{3}%
{2}-n;1\right)\nonumber
\end{eqnarray}
{where }$_{3}F_{2}$ is the generalized hypergeometric function. This expression
can be worked under different assumptions to end up with a simple form and
therefore make link with some physical phenomena. Let us illustrate \eqref{cccc} by
summarizing some simple forms in Table 2

\begin{center}%
\begin{tabular}
[c]{|c|c|c|}\hline\hline
Configuration & Quantum numbers $n,l,s$ & Angular current $\bar{J}_{\varphi
}^{n,l,s}$\\\hline\hline
$1$ & $n=0,l=0,s=+1$ & $\frac{1}{4}\left(  1+3\kappa\right)  \sqrt{\frac
{\pi\left\vert G\right\vert }{m^{2}+4\left\vert G\right\vert }}$\\\hline\hline
$2$ & $n=0,l=0,s=-1$ & $-\frac{1}{2}\left(  1+\kappa\right)  \sqrt{\frac
{\pi\left\vert G\right\vert }{m^{2}+2\left\vert G\right\vert \left(
1+\kappa\right)  }}$\\\hline\hline
$3$ & $n=0,l=1,s=+1$ & $\frac{1}{4}\left(  1-3\kappa\right)  \sqrt{\frac
{\pi\left\vert G\right\vert }{m^{2}+4\left\vert G\right\vert }}$\\\hline\hline
$4$ & $n=0,l=1,s=-1$ & $-\frac{1}{2}\left(  1-\kappa\right)  \sqrt{\frac
{\pi\left\vert G\right\vert }{m^{2}+2\left\vert G\right\vert \left(
1-\kappa\right)  }}$\\\hline\hline
\end{tabular}
\end{center}
Table 2: {\sf Table illustrates some examples of the angular current for
some quantum numbers values }$(n,l,s)$.\\

From Table 2, we can immediately realize that there are some symmetries.
Indeed, for the configurations $(1,3)$ one can establish the relation%
\begin{equation}
\bar{J}_{\varphi}^{0,0,+1}\left(  \kappa\right)  =\bar{J}_{\varphi}%
^{0,1,+1}\left(  -\kappa\right)
\end{equation}
and the same between $(2,4)$%
\begin{equation}
\bar{J}_{\varphi}^{0,0,-1}\left(  \kappa\right)  =\bar{J}_{\varphi}%
^{0,1,-1}\left(  -\kappa\right)
\end{equation}
under the change $\kappa\longrightarrow-\kappa$, with the parameter
$\kappa=\mbox{sgn}\left(  G\right)  =\frac{G}{\left\vert G\right\vert }\neq0$.
We expect to have other symmetries can be found by choosing new
configurations of the three quantum numbers. This is interesting and could be
used to formulate a theory to describe low dimensional systems without having
external excitations.

%%%%%%%%%%%%%%%%%%%%%%%%%%%%%%%%%%%%%%%%%
\section{Nonrelativistic limit}
%%%%%%%%%%%%%%%%%%%%%%%%%%%%%%%%%%%%%%%%%%

To recover the nonrelativistic limit, we use the standard process. Indeed, by
requiring the limit $m\longrightarrow\infty$ in the obtained solutions of the
energy spectrum, one can see%
\begin{equation}
\mathcal{E}_{n,l,s}\longrightarrow sm\left(  1+\frac{\left\vert G\right\vert
}{m^{2}}\left[  2n+1+\left\vert l-\tfrac{s+1}{2}\right\vert -\tfrac
{G}{\left\vert G\right\vert }\left(  l+\tfrac{s-1}{2}\right)  \right]
\right)
\end{equation}
\begin{equation}
\sqrt{\tfrac{\mathcal{E}_{n,l,s}+sm}{2\mathcal{E}_{n,l,s}}}\longrightarrow1
\end{equation}
\begin{equation}
\sqrt{\tfrac{\mathcal{E}_{n,l,s}-sm}{2\mathcal{E}_{n,l,s}}}\longrightarrow0.
\end{equation}
These allow the Green function to behave as
\begin{eqnarray}
S^{c}(\mathbf{x}_{b},\mathbf{x}_{a}) &  \longrightarrow & i\tfrac{\left\vert
G\right\vert }{\pi}\sum_{n=0}^{+\infty}\sum_{l=-\infty}^{+\infty}\sum_{s=\pm
1}e^{i(l-(1+s)/2)\left(  \varphi_{b}-\varphi_{a}\right)  }e^{-i\left(
sm+s\frac{\left\vert G\right\vert }{m}\left[  2n+1+\left\vert l-\tfrac{s+1}%
{2}\right\vert -\tfrac{G}{\left\vert G\right\vert }\left(  l+\tfrac{s-1}%
{2}\right)  \right]  \right)  \left(  t_{b}-t_{a}\right)  } \nonumber\\
&& \times \Theta\left(
s\left(  t_{b}-t_{a}\right)  \right)
 F_{n,l,s}^{\kappa^{\prime}}\left(  z_{b}\right)  F_{n,l,s}%
^{\kappa^{\prime}}\left(  z_{a}\right)  \chi_{s}\chi_{s}^{+} \label{95}.%
\end{eqnarray}
Now taking $l\longrightarrow l+\frac{s+1}{2}$ to write \eqref{95} as
\begin{eqnarray}
S^{c}(\mathbf{x}_{b},\mathbf{x}_{a}) &  \longrightarrow & i\tfrac{\left\vert
G\right\vert }{\pi}\sum_{n=0}^{+\infty}\sum_{l=-\infty}^{+\infty}e^{il\left(
\varphi_{b}-\varphi_{a}\right)  }e^{-i\left(  m+\frac{\left\vert G\right\vert
}{m}\left[  2n+1+\left\vert l\right\vert -\tfrac{G}{\left\vert G\right\vert
}\left(  l+1\right)  \right]  \right)  \left(  t_{b}-t_{a}\right)  }%
\Theta\left(  \left(  t_{b}-t_{a}\right)  \right)  \nonumber\\
&&  \times F_{n,l}\left(  z_{b}\right)  F_{n,l,}\left(  z_{a}\right)  \chi
_{+1}\chi_{+1}^{+}\\
&&  +i\tfrac{\left\vert G\right\vert }{\pi}\sum_{n=0}^{+\infty}\sum_{l=-\infty
}^{+\infty}e^{il\left(  \varphi_{b}-\varphi_{a}\right)  }e^{i\left(
m+\frac{\left\vert G\right\vert }{m}\left[  2n+1+\left\vert l\right\vert
-\tfrac{G}{\left\vert G\right\vert }\left(  l-1\right)  \right]  \right)
\left(  t_{b}-t_{a}\right)  }\Theta\left(  -\left(  t_{b}-t_{a}\right)
\right)  \nonumber\\
&&  \times F_{n,l}\left(  z_{b}\right)  F_{n,l}\left(  z_{a}\right)  \chi
_{-1}\chi_{-1}^{+}\nonumber %
\end{eqnarray}
and therefore to end up with usual eigenfunctions for the nonrelativistic
problem
\begin{equation}
\Psi_{n,l,s}\left(  r,\varphi;t\right)  \rightarrow e^{-i\left(
m+\frac{\left\vert G\right\vert }{m}\left[  2n+1+\left\vert l\right\vert
-\tfrac{G}{\left\vert G\right\vert }\left(  l+1\right)  \right]  \right)
t}\left(
\begin{array}
[c]{c}%
\Psi^{NR}\left(  r,\varphi\right)  \\
0
\end{array}
\right)  \qquad\text{ }%
\end{equation}
or in the form
\begin{equation}
\Psi_{n,l,s}\left(  r,\varphi;t\right)  \rightarrow e^{+i\left(
m+\frac{\left\vert G\right\vert }{m}\left[  2n+1+\left\vert l\right\vert
-\tfrac{G}{\left\vert G\right\vert }\left(  l-1\right)  \right]  \right)
t}\left(
\begin{array}
[c]{c}%
0\\
\Psi^{NR}\left(  r,\varphi\right)
\end{array}
\right)
\end{equation}
where $\Psi^{NR}\left(r,\varphi\right)  $ are the nonrelativistic functions
defined by%
\begin{equation}
\Psi^{NR}\left(  r,\varphi\right)  =\sqrt{\tfrac{\left\vert G\right\vert
n!}{\pi\left(  n+\left\vert l\right\vert \right)  !}}e^{il\varphi}\left(
\sqrt{\left\vert G\right\vert }r\right)  ^{\left\vert l\right\vert }%
e^{-\frac{1}{2}\left\vert G\right\vert r^{2}}L_{n}^{\left\vert l\right\vert
}\left(  \left\vert G\right\vert r^{2}\right)
\end{equation}
and the corresponding nonrelativistic energy is
\begin{equation}
\mathcal{E}^{NR}=\frac{\left\vert G\right\vert }{m}\left[  2n+1+\left\vert
l\right\vert -\frac{G}{\left\vert G\right\vert }\left(  l\pm1\right)  \right]
=\frac{\mathcal{E}_{n,l,s}^{2}-m^{2}}{2m}.%
\end{equation}

The same results can be found for one-relativistic particle living on the
plane $(x;y)$ in presence of a perpendicular magnetic field $B$ described by
the Pauli-Schr\"{o}dinger Hamiltonian%
\begin{equation}
H^{PS}=\frac{1}{2m}\left[  \vec{\sigma}.\left(  \vec{P}-e\vec{A}\right)
\right]  ^{2}%
\end{equation}
in the symmetric gauge $\vec{A}=\frac{B}{2}\left(  -y,x\right)  $, more detail
can be found in \cite{6}. This tells us that our findings are interesting and
allow to recover the well-known results for the present problem.

%%%%%%%%%%%%%%%%%%%%%%%%%%%%%
\section{Completeness}
%%%%%%%%%%%%%%%%%%%%%%%%%%%%%%

It is interesting to examine  the completeness relation associated to our solutions
of the energy spectrum. Indeed,
the closure relation obeyed by eigenfunctions is given by
\begin{equation}
\sum_{n=0}^{+\infty}\sum_{l=-\infty}^{+\infty}\sum_{s=\pm1}\Psi_{n,l,s}%
^{\kappa,\kappa^{\prime}}\left(  \mathbf{x}_{b}\right)  \left(  \Psi
_{n,l,s}^{\kappa,\kappa^{\prime}}\left(  \mathbf{x}_{a}\right)  \right)
^{+}=\mathbb{I}_{2\times2}\delta\left(  \mathbf{x}_{b}-\mathbf{x}_{a}\right).
\end{equation}
Introducing%
\begin{eqnarray}
I\left(  \mathbf{x}_{b},\mathbf{x}_{a}\right)   &  =&\sum_{n=0}^{+\infty}%
\sum_{l=-\infty}^{+\infty}\sum_{s=\pm1}\Psi_{n,l,s}^{\kappa,\kappa^{\prime}%
}\left(  r_{b},\varphi_{b}\right)  \left(  \Psi_{n,l,s}^{\kappa,\kappa
^{\prime}}\left(  r_{a},\varphi_{a}\right)  \right)  ^{+}\nonumber\\
&  =&\sqrt{\tfrac{\left\vert G\right\vert }{\pi}}e^{il\varphi_{b}}\left[
\sqrt{\tfrac{\mathcal{E}_{n,l,s}+sm}{2\mathcal{E}_{n,l,s}}}F_{n,l,s}%
^{\kappa^{\prime}}\left(  z\right)  \chi_{s}-s\sqrt{\tfrac{\mathcal{E}%
_{n,l,s}-sm}{2\mathcal{E}_{n,l,s}}}F_{n-\frac{\kappa^{\prime}+\kappa}%
{2}s,l,-s}^{\kappa^{\prime}}\left(  z\right)  \chi_{-s}\right] \nonumber\\
&&  \times\sqrt{\tfrac{\left\vert G\right\vert }{\pi}}\left[  \sqrt
{\tfrac{\mathcal{E}_{n,l,s}+sm}{2\mathcal{E}_{n,l,s}}}F_{n,l,s}^{\kappa
^{\prime}}\left(  z\right)  \chi_{s}^{+}-s\sqrt{\tfrac{\mathcal{E}_{n,l,s}%
-sm}{2\mathcal{E}_{n,l,s}}}F_{n-\frac{\kappa^{\prime}+\kappa}{2}%
s,l,-s}^{\kappa^{\prime}}\left(  z\right)  \chi_{-s}^{+}\right] \nonumber\\
&  =& \tfrac{\left\vert G\right\vert }{\pi}\sum_{n=0}^{+\infty}\sum_{l=-\infty
}^{+\infty}\sum_{s=\pm1}\left\{  \tfrac{\mathcal{E}_{n,l,s}+sm}{2\mathcal{E}%
_{n,l,s}}F_{n,l,s}^{\kappa^{\prime}}\left(  z_{b}\right)  F_{n,l,s}%
^{\kappa^{\prime}}\left(  z_{a}\right)  e^{-i\frac{s+1}{2}\varphi_{b}%
}e^{i\frac{s+1}{2}\varphi_{a}}\chi_{s}\chi_{s}^{+}\right. \nonumber\\
&&  +\tfrac{\mathcal{E}_{n,l,s}-sm}{2\mathcal{E}_{n,l,s}}F_{n-\frac
{\kappa^{\prime}+\kappa}{2}s,l,-s}^{\kappa^{\prime}}\left(  z_{b}\right)
F_{n-\frac{\kappa^{\prime}+\kappa}{2}s,l,-s}^{\kappa^{\prime}}\left(
z_{a}\right)  e^{-i\frac{-s+1}{2}\varphi_{b}}e^{i\frac{-s+1}{2}\varphi_{a}%
}\chi_{-s}\chi_{-s}^{+}\\
& & -s\tfrac{\mathcal{E}_{n,l,s}-sm}{2\mathcal{E}_{n,l,s}}\tfrac{\mathcal{E}%
_{n,l,s}+sm}{2\mathcal{E}_{n,l,s}}F_{n-\frac{\kappa^{\prime}+\kappa}{2}%
s,l,-s}^{\kappa^{\prime}}\left(  z_{b}\right)  F_{n,l,s}^{\kappa^{\prime}%
}\left(  z_{a}\right)  e^{-i\frac{-s+1}{2}\varphi_{b}}e^{i\frac{s+1}{2}%
\varphi_{a}}\chi_{-s}\chi_{s}^{+}\nonumber\\
&&  \left.  -s\tfrac{\mathcal{E}_{n,l,s}-sm}{2\mathcal{E}_{n,l,s}}%
\tfrac{\mathcal{E}_{n,l,s}+sm}{2\mathcal{E}_{n,l,s}}F_{n-\frac{\kappa^{\prime
}+\kappa}{2}s,l,-s}^{\kappa^{\prime}}\left(  z_{a}\right)  F_{n,l,s}%
^{\kappa^{\prime}}\left(  z_{b}\right)  e^{-i\frac{s+1}{2}\varphi_{b}%
}e^{i\frac{-s+1}{2}\varphi_{a}}\chi_{s}\chi_{-s}^{+}\right\}\nonumber.
\end{eqnarray}
Taking into account of
the symmetry properties discussed in section 5, in particular %in Table 1 and specifically   obtained for the eigensolutions
%according to third line of Table 1, i.e.
the changes $
s\rightarrow-s$ and $n\rightarrow n-\frac{\kappa^{\prime}+\kappa}{2}s$,
one can simplify $I\left(  \mathbf{x}_{b},\mathbf{x}_{a}\right) $ as
\begin{eqnarray}
I\left(  \mathbf{x}_{b},\mathbf{x}_{a}\right)   &  =& \tfrac{\left\vert
G\right\vert }{\pi}\sum_{n=0}^{+\infty}\sum_{l=-\infty}^{+\infty}\sum_{s=\pm
1}F_{n,l,s}^{\kappa^{\prime}}\left(  z_{b}\right)  F_{n,l,s}^{\kappa^{\prime}%
}\left(  z_{a}\right)  e^{-i\frac{s+1}{2}\varphi_{b}}e^{i\frac{s+1}{2}%
\varphi_{a}}\chi_{s}\chi_{s}^{+}\nonumber\\
& =& \tfrac{\left\vert G\right\vert }{\pi}\sum_{n=0}^{+\infty}\sum_{l=-\infty
}^{+\infty}\sum_{s=\pm1}\left\{  \tfrac{\left(  n\right)  !}{\left(
n+\kappa^{\prime}\left(  l-\frac{s+1}{2}\right)  \right)  !}e^{-\frac{1}%
{2}z_{b}^{2}}z_{b}^{\kappa^{\prime}\left(  l-\frac{s+1}{2}\right)  }%
L_{n}^{\kappa^{\prime}\left(  l-\frac{s+1}{2}\right)  }\left(  z_{b}%
^{2}\right)  \right. \\
&&  \times \left.  e^{-\frac{1}{2}z_{a}^{2}}z_{a}^{\kappa^{\prime}\left(  l-\frac
{s+1}{2}\right)  }L_{n}^{\kappa^{\prime}\left(  l-\frac{s+1}{2}\right)
}\left(  z_{a}^{2}\right)  e^{-i\frac{s+1}{2}\varphi_{b}}e^{i\frac{s+1}%
{2}\varphi_{a}}\chi_{s}\chi_{s}^{+}\right\}\nonumber.
\end{eqnarray}
Taking $l\rightarrow l+\frac{s+1}{2}$, $I\left(  \mathbf{x}_{b}%
,\mathbf{x}_{a}\right)  $ becomes%
\begin{equation}
I\left(  \mathbf{x}_{b},\mathbf{x}_{a}\right)  =2\left\vert G\right\vert
\sum_{n=0}^{+\infty}\tfrac{\left(  n\right)  !}{\left(  n+\kappa^{\prime
}l\right)  !}e^{-\frac{1}{2}z_{b}^{2}}z_{b}^{\kappa^{\prime}l}L_{n}%
^{\kappa^{\prime}l}\left(  z_{b}^{2}\right)  e^{-\frac{1}{2}z_{a}^{2}}%
z_{a}^{\kappa^{\prime}l}L_{n}^{\kappa^{\prime}l}\left(  z_{a}^{2}\right)
\sum_{l=-\infty}^{+\infty}\frac{e^{il\left(  \varphi_{b}-\varphi_{a}\right)
}}{2\pi}\sum_{s=\pm1}\chi_{s}\chi_{s}^{+}.%
\end{equation}
By making use of the known closure relation obeyed by the generalized Laguerre
functions%
\begin{equation}
\sum_{n=0}^{+\infty}\tfrac{\left(  n\right)  !}{\left(  n+\alpha\right)
!}e^{-\frac{1}{2}\left(  \rho_{a}+\rho_{b}\right)  }\rho_{a}^{\frac{\alpha}%
{2}}\rho_{b}^{\frac{\alpha}{2}}L_{n}^{\alpha}\left(  \rho_{a}\right)
L_{n}^{\alpha}\left(  \rho_{b}\right)  =\delta\left(  \rho_{b}-\rho
_{a}\right)
\end{equation}
to obtain
\begin{align}
I\left(  \mathbf{x}_{b},\mathbf{x}_{a}\right)   &  =2\left\vert G\right\vert
\sum_{n=0}^{+\infty}\tfrac{\left(  n\right)  !}{\left(  n+\kappa^{\prime
}l\right)  !}e^{-\frac{1}{2}z_{b}^{2}}z_{b}^{\kappa^{\prime}l}L_{n}%
^{\kappa^{\prime}l}\left(  z_{b}^{2}\right)  e^{-\frac{1}{2}z_{a}^{2}}%
z_{a}^{\kappa^{\prime}l}L_{n}^{\kappa^{\prime}l}\left(  z_{a}^{2}\right)
\sum_{l=-\infty}^{+\infty}\frac{e^{il\left(  \varphi_{b}-\varphi_{a}\right)
}}{2\pi}\sum_{s=\pm1}\chi_{s}\chi_{s}^{+}\nonumber\\
&  =2\left\vert G\right\vert \delta\left(  z_{b}^{2}-z_{a}^{2}\right)
\delta\left(  \varphi_{b}-\varphi_{a}\right)  \mathbb{I}_{2\times2}=2\left\vert
G\right\vert \delta\left(  \left\vert G\right\vert r_{b}^{2}-\left\vert
G\right\vert r_{a}^{2}\right)  \delta\left(  \varphi_{b}-\varphi_{a}\right)
\mathbb{I}_{2\times2}%
\end{align}
Now considering the properties of the Dirac delta function
\begin{equation}
\delta\left(  \left\vert G\right\vert r_{b}^{2}-\left\vert G\right\vert
r_{a}^{2}\right)  =\frac{\delta\left(  r_{b}-r_{a}\right)  }{2\left\vert
G\right\vert \sqrt{r_{b}r_{a}}}%
\end{equation}
to finally get
\begin{equation}
I\left(  \mathbf{x}_{b},\mathbf{x}_{a}\right)  =\frac{\delta\left(
r_{b}-r_{a}\right)  }{\sqrt{r_{b}r_{a}}}\delta\left(  \varphi_{b}-\varphi
_{a}\right)  \mathbb{I}_{2\times2}=\mathbb{I}_{2\times2}\delta\left(  \mathbf{x}_{b}%
-\mathbf{x}_{a}\right)
\end{equation}
which proves the completeness of the Dirac oscillator eigenfunctions obtained above.

We close by noting that
choosing
%Let us summarize by noting that in our case the choice of
the energy sign equal to that
of the quantum number $s$, i.e.
%notons que dans notre cas le fet de choisir le signe de l'energie est celui du
%nombre quantique $s$  tel que
$\mbox{sgn}\left(  \mathcal{E}_{n,l,s}%
\right)  =\mbox{sgn}\left(  s\right)  $, is required by the supersymmetric nature of our spinors
solution of the Dirac equation. This together with the quantification of the radial variable suggest to
take the quantum number $n$ belongs to $\mathbb{N}$.
%% ce choix est naturel et fait partie a
%la nature supersymetrique des spineures de l'equation de Dirac, the range of
%$n$ is $\left[  0,+\infty\right]  $ est naturel aussi a cause de la
%quantification de la variable radial $r.$
However, in \cite{24} the relation $\mbox{sgn}\left(\mathcal{E}_{n,l,s}\right)  =\mbox{sgn}\left(n\right)$ is made by hand
and therefore $n$ must be in $\mathbb{Z}$ to prove the
completeness of the Dirac oscillator in 3-dimensional.

%%%%%%%%%%%%%%%%%%%%%%%%%%%%%%%%%
\section{Conclusion}
%%%%%%%%%%%%%%%%%%%%%%%%%%%%%%%%%%

We have established a new symmetric expression to solve the problem of the
relativistic confining fermion interacting with the constant magnetic field
within the framework supersymmetric representation path integrals of Fradkin
and Gitman \cite{19}. This new symmetric form for global path integral
represent the first attempt to find a general method to extract spinors in
certain elegance. The energy spectrum is derived from the spectral
decomposition of the causal Green function and have found to be in agreement
with that obtained \cite{15}.

The solutions of the energy spectrum are obtained to be dependent on different
physical parameters and quantum numbers. By inspecting the basic features of
such solutions, we have showed there our system is hidden some interesting
symmetries. Indeed, by considering three different configurations of the
quantum numbers, we have noticed that the energy remained invariant. This
properties could be used to systemically establish a theory based on such
symmetries to deal with some physical phenomena like the quantum Hall
effect \cite{14}.

Subsequently, we have focused on two important issues related to our system.
The first one is the density of current where the radial part was found to be
null, while the angular one was not and was expressed in terms of the
generalized hypergeometric function. The second issue is the nonrelativistic
limit that has been studied by using the standard method and therefore the
corresponding solutions of the energy spectrum were recovered in accordance
with \cite{15}.

To close, let us notice that the advantage of using the path integral techniques for the
system under consideration is to extract  the corresponding spinors
%considering our problem that the spinors that we extract
from the spectral decomposition
of the Green function in the simple and compact forms compared to those obtained in \cite{15}. These forms make them suitable for variational calculations
such that
the currant density, the nonrelativistic
limit and  etc.

%%%%%%%%%%%%%%%%%%%%%%%%%%%%%%%%%%%%%%%%%
\section*{Acknowledgments}
%%%%%%%%%%%%%%%%%%%%%%%%%%%%%%%%%%%%%%%

A.M and A.J acknowledge the financial support from King Faisal University. The
present work was done under Project Number 140233, `Path Integral Techniques
for Interacting Dirac Particles'. A.M and A.J are very grateful to  T.
Sbeouelji for numerous helpful discussions and assistant.


\begin{thebibliography}{99}                                                                                               %


\bibitem {1}A.M.J. Schakel, Phys. Rev. D43 (1991) 1428; A. Neagu and A.M.J.
Schakel, Phys. Rev. D48(1993) 1785.

\bibitem {2}M.O. Goerbig and N. Regnault, Phys. Rev. B74 (2006 ) 161407.

\bibitem {3}C. T\"{o}ke, P.E. Lammert, J.K. Jain and V.H. Crespi, Phys. Rev.
B74 (2006) 235417.

\bibitem {4}D.V. Khveshchenko, Phys. Rev. B75 (2007) 153405.

\bibitem {5}C. Toke and J.K. Jain, Phys. Rev. B75 (2007) 244540.

\bibitem {6}A. Jellal, Nucl. Phys. B804 (2008) 361.

\bibitem {7}J. Karwowski and G. Pestka, Theoretical Chemistry Accounts 118
(2007) 519.

\bibitem {8}P.A. Cook, Lettere al Nuovo Cimento 1 (1971) 419.

\bibitem {9}M. Moshinsky and A. Szczepaniak, J. Phys. A: Math. Gen. 22 (1989) L817-82.

\bibitem {10}A.K. Geim and K. S. Novoselov, Nat. Mater. 6 (2007) 183.

\bibitem {11}K.S. Novoselov, A.K. Geim, S.V. Morozov, D. Jiang, Y. Zhang, S.V.
Dubonos, I.V. Grigorieva and A.A. Firsov, Science 306 (2004) 666.

\bibitem {12}A.H. Castro Neto, F. Guinea, N.M.R. Peres, K.S. Novoselov and
A.K. Geim, Rev. Mod. Phys. 81(2009) 109.

\bibitem {13}M. Grundmann, F. Heinrichsdorff, C. Ribbat, M.-H. Mao and D.
Bimberg, Applied Physics:Lasers and Optics B69 (1999) 413; T. Pohjola, D.
Boese, H. Schoeller, J. K nig and G. Sch n,Physica: Condensed Matter B284-288
(2000) 1762; D. Vanmaekelbergh and P. Liljeroth, Chem.Soc. Rev. 34 (2005) 299;
J.L. West and N.J. Halas, Annu. Rev. Biomed. Eng. 5 (2003) 285; H.Arya, Z.
Kaul, R. Wadhwa, K. Taira, T. Hirano and S.C. Kaul, Biochem. Biophys. Res.
Commun.329 (2005) 1173; E.J. Gansen, M.A. Rowe, M.B. Greene, D. Rosenberg,
T.E. Harvey, M.Y. Su,R.H. Had eld, S.W. Nam and R.P.Mirin, Nature Photonics 1
(2007) 585.

\bibitem {14}R.E. Prange and S.M. Girvin, editors, "The Quantum Hall Effect",
(Springer, New York 1990).

\bibitem {15}A. Jellal, A.D. Alhaidari and H. Bahlouli, Phys. Rev A80 (2009) 012109.

\bibitem {16}V.M. Villalba and A.A.R. Maggiolo, Eur. J. Phys. B22 (2001) 31.

\bibitem {17}P.R. Holland, Foundations of Physics 16 (1986) 701.

\bibitem {18}C. Alexandrou, R. Rosenfelder, and A.W. Schreiber, Phys. Rev. A59
(1998) 3.

\bibitem {19}E.S. Fradkin and D.M. Gitman, Phys. Rev. D44 (1991) 3230; S.
Haouat and L. Chetouani, Z. Naturforsch 62a (2007) 34.

\bibitem {20}N. Boudiaf, A. Merdaci and L. Chetouani, J. Phys. A: Math. Theor.
42 (2009)  015303; A. Merdaci, N. Boudiaf and L. Chetouani, Z. Naturforsch. 63a (2008) 283.

\bibitem {21}D.M. Gitman and S.I. Zlatev, Phys. Rev. D55 (1997) 7701.

\bibitem {22}I.S. Gradshteyn and I.M. Ryzhik, "Table of Integrals, Series, And
Products" (Academic Press,New York 1980).

\bibitem {233} O. Yilmaz, M. Saglam and Z. Z. Aydin, Concepts of Physics 4 (2007) 141.

\bibitem {23}H. Benzair, M. Merad, T. Boudjedaa and A. Makhlouf, Z.
Naturforsch 67a (2012) 77.

\bibitem {24}Rados\l aw Szmytkowski and Marek Gruchowski, J. Phys. A: Math.
Gen. 34 (2001) 4991.
\end{thebibliography}
\end{document}